\documentstyle[12pt,aasms4]{article}

\newcommand{\arcsecpoint}{\ifmmode ''\!. \else $''\!.$\fi}

\newcommand{\kms}{\ifmmode {\rm km\ s}^{-1} \else km s$^{-1}$\fi}
\newcommand{\Msun}{\ifmmode {\rm M}_{\odot} \else M$_{\odot}$\fi}
\newcommand{\Lsun}{\ifmmode {\rm L}_{\odot} \else L$_{\odot}$\fi}
\newcommand{\qo}{\ifmmode q_{\rm o} \else $q_{\rm o}$\fi}
\newcommand{\Ho}{\ifmmode H_{\rm o} \else $H_{\rm o}$\fi}
\newcommand{\ho}{\ifmmode h_{\rm o} \else $h_{\rm o}$\fi}

\newcommand{\vFWHM}{\ifmmode v_{\mbox{\tiny FWHM}} \else
                    $v_{\mbox{\tiny FWHM}}$\fi}
\newcommand{\CCF}{\ifmmode F_{\it CCF} \else $F_{\it CCF}$\fi}
\newcommand{\ACF}{\ifmmode F_{\it ACF} \else $F_{\it ACF}$\fi}
\newcommand{\Halpha}{\ifmmode {\rm H}\alpha \else H$\alpha$\fi}
\newcommand{\Hbeta}{\ifmmode {\rm H}\beta \else H$\beta$\fi}
\newcommand{\Hgamma}{\ifmmode {\rm H}\gamma \else H$\gamma$\fi}
\newcommand{\Hdelta}{\ifmmode {\rm H}\delta \else H$\delta$\fi}
\newcommand{\Lya}{\ifmmode {\rm Ly}\alpha \else Ly$\alpha$\fi}
\newcommand{\Lyb}{\ifmmode {\rm Ly}\beta \else Ly$\beta$\fi}
\newcommand{\HeI}{\ifmmode {\rm He}\,{\sc i}\,\lambda5876 \else 
	          He\,{\sc i}\,$\lambda5876$\fi}
\newcommand{\HeII}{\ifmmode {\rm He}\,{\sc ii}\,\lambda4686 \else 
	           He\,{\sc ii}\,$\lambda4686$\fi}

\newcommand{\ciii}{\ifmmode {\rm C}\,{\sc iii} \else C\,{\sc iii}\fi}
\newcommand{\civ}{\ifmmode {\rm C}\,{\sc iv} \else C\,{\sc iv}\fi}
\newcommand{\nii}{N\,{\sc ii}}

\newcommand{\oi}{O\,{\sc i}}

\newcommand{\oiii}{O\,{\sc iii}}
\newcommand{\ob}{[O\,{\sc iii}]\,$\lambda5007$}

\newcommand{\feii}{Fe\,{\sc ii}}

\begin{document}

\slugcomment{To be published in Astrophys.J.Suppl. 115, April 1998}

%
%

\title{STEPS TOWARD DETERMINATION OF THE SIZE AND STRUCTURE OF THE 
       BROAD-LINE REGION IN ACTIVE GALACTIC NUCLEI. 
       XII. GROUND-BASED MONITORING OF 3C\,390.3}

\author{M.\,Dietrich
\altaffilmark{1}
B.\,M.\,Peterson
\altaffilmark{2}
P.\,Albrecht
\altaffilmark{3}
M.\,Altmann
\altaffilmark{4}
A.\,J.\,Barth
\altaffilmark{5}
P.\,J.\,Bennie 
\altaffilmark{6}
R.\,Bertram
\altaffilmark{2,7}
N.\,G.\,Bochkarev
\altaffilmark{8}
H.\,Bock
\altaffilmark{1}
J.\,M.\,Braun
\altaffilmark{4}
A.\,Burenkov
\altaffilmark{9}
S.\,Collier
\altaffilmark{6}
L.-Z.\,Fang
\altaffilmark{10}
O.\,P.\,Francis
\altaffilmark{11}
A.\,V.\,Filippenko
\altaffilmark{5}
C.\,B.\,Foltz
\altaffilmark{12}
W.\,G\"{a}ssler
\altaffilmark{1,13}
C.\,M.\,Gaskell
\altaffilmark{11}
M.\,Geffert
\altaffilmark{4}
K.\,K.\,Ghosh
\altaffilmark{14}
R.\,W.\,Hilditch
\altaffilmark{6}
R.\,K.\,Honeycutt
\altaffilmark{15}
K.\,Horne
\altaffilmark{6}
J.\,P.\,Huchra
\altaffilmark{16}
S.\,Kaspi
\altaffilmark{17}
M.\,K\"{u}mmel
\altaffilmark{1}
K.\,M.\,Leighly
\altaffilmark{18}
D.\,C.\,Leonard
\altaffilmark{5}
Yu.F.\,Malkov
\altaffilmark{19}
V.\,Mikhailov
\altaffilmark{9}
H.\,R.\,Miller
\altaffilmark{20}
A.\,C.\,Morrill
\altaffilmark{21}
J.\,Noble
\altaffilmark{22}
P.\,T.\,O'Brien
\altaffilmark{23}
T.\,D.\,Oswalt
\altaffilmark{24}
S.\,P.\,Pebley
\altaffilmark{11}
M.\,Pfeiffer
\altaffilmark{1}
V.\,I.\,Pronik
\altaffilmark{19}
B.-C.\,Qian
\altaffilmark{25}
J.\,W.\,Robertson
\altaffilmark{15}
A.\,Robinson
\altaffilmark{26}
K.\,S.\,Rumstay
\altaffilmark{27}
J.\,Schmoll
\altaffilmark{4,28}
S.\,G.\,Sergeev
\altaffilmark{19}
E.\,A.\,Sergeeva
\altaffilmark{19}
A.\,I.\,Shapovalova
\altaffilmark{9}
D.\,R.\,Skillman
\altaffilmark{29}
S.\,A.\,Snedden
\altaffilmark{11}
S.\,Soundararajaperumal
\altaffilmark{14}
G.\,M.\,Stirpe
\altaffilmark{30}
J.\,Tao
\altaffilmark{24}
G.\,W.\,Turner
\altaffilmark{15}
R.\,M.\,Wagner
\altaffilmark{2,7}
S.\,J.\,Wagner
\altaffilmark{1}
J.\,Y.\,Wei
\altaffilmark{31}
H.\,Wu
\altaffilmark{31}
W.\,Zheng
\altaffilmark{32}
\and
Z.\,L.\,Zou
\altaffilmark{31}
}
\altaffiltext{1}
{Landessternwarte Heidelberg, K\"{o}nigstuhl, D-69117 Heidelberg, Germany}
\altaffiltext{2}
{Department of Astronomy, Ohio State University, 174 West 18th Ave., 
Columbus OH 43210-1106}
\altaffiltext{3}
{Universit\"{a}tssternwarte G\"{o}ttingen, Geismarlandstra\ss e 11,
 D-37083 G\"{o}ttingen, Germany}
\altaffiltext{4}
{Sternwarte der Universit\"{a}t Bonn, Auf dem H\"{u}gel 71, D-53121 Bonn, 
 Germany}
\altaffiltext{5}
{Department of Astronomy, University of California, 601 Campbell Hall, 
 Berkeley, CA 94720-3411}
\altaffiltext{6}
{University of St.\ Andrews, School of Physics and Astronomy, North Haugh,
 St.\ Andrews, Fife KY16 9SS, Scotland, UK}
\altaffiltext{7}
{Mailing address: Lowell Observatory, 1400 Mars Hill Road, 
 Flagstaff, AZ 86001}
\altaffiltext{8}
{Sternberg Astronomical Institute, Universitetskij Prospect, 13, 
 119899 Moscow, Russia}
\altaffiltext{9}
{Special Astrophysical Observatory, Russian Academy of Science,
 Nyzknij Arkhyz, Karachaj-Cherkess Republic, 357147, Russia}
\altaffiltext{10}
{Department of Physics, University of Arizona, Tucson, AZ 85721}
\altaffiltext{11}
{University of Nebraska, Lincoln, Department of Physics and Astronomy,
 Lincoln, NE 68588-0111}
\altaffiltext{12}
{MMT Observatory, University of Arizona, Tucson, AZ 85721}
\altaffiltext{13}
{Universit\"{a}tssternwarte M\"{u}nchen, Scheinerstr. 1,
 D-81679 M\"{u}nchen, Germany}
\altaffiltext{14}
{Indian Institute of Astrophysics, Vainu Bappu Observatory, Kavalur, 
 Alangayam, N.A.A.\ 635 701, T.N., India}
\altaffiltext{15}
{Department of Astronomy, Indiana University, 319 Swain West, Bloomington,
 IN 47405}
\altaffiltext{16}
{Harvard-Smithsonian Center for Astrophysics, 60 Garden Street, 
 Cambridge, MA 02138}
\altaffiltext{17}
{Department of Physics and Astronomy, Tel Aviv University, Tel Aviv 69978,
 Israel}
\altaffiltext{18}
{Columbia Astrophysics Laboratory, Columbia University, 
 538 West 120$^{\rm th}$ Street, New York, NY 10027}
\altaffiltext{19}
{Crimean Astrophysical Observatory, p/o Nauchny, 334413 Crimea, Ukraine}
\altaffiltext{20}
{Department of Physics and Astronomy, Georgia State University, Atlanta, 
 GA 30303}
\altaffiltext{21}
{Center for Space Physics, Boston University, 725 Commonwealth Ave,
 Boston, MA 02215}
\altaffiltext{22}
{Center for Automated Space Sciences, Dept. of Physics and Astronomy,
 Western Kentucky University,Bowling Green, KY 42101-3576}
\altaffiltext{23}
{Department of Physics \& Astronomy, University of Leicester,
 University Road, Leicester LE1 7RH, UK}
\altaffiltext{24}
{Dept.\ of Physics and Space Science,  Florida Institute of Technology, 
 150 West University Avenue, Melbourne, FL 32901-6988}
\altaffiltext{25}
{Shanghai Observatory, The Chinese Academy of Sciences, 80 Nandam Road,
 Shanghai 200030, China}
\altaffiltext{26}
{Division of Physics and Astronomy, Dept.\ of Physical Science, University 
 of Hertfordshire, College Lane, Hatfield, Herts AL109AB, UK}
\altaffiltext{27}
{Department of Physics, Astronomy, and Geology, Valdosta State University,
 Valdosta, GA 31698-0055}
\altaffiltext{28}
{Astrophysikalisches Institut Potsdam, An der Sternwarte 16,
 D-14482 Potsdam, Germany}
\altaffiltext{29}
{Center for Basement Astrophysics, 9517 Washington Avenue, Laurel, MD 20723}
\altaffiltext{30}
{Osservatorio Astronomica di Bologna, Via Zamboni 33, I-40126 Bologna, Italy}
\altaffiltext{31}
{Beijing Astronomical Observatory, Beijing 100080, Beijing, China}
\altaffiltext{32}
{Center for Astrophysical Sciences, John Hopkins University. Baltimore, 
MD 21218}
%
\begin{abstract}
Results of a ground-based optical monitoring campaign on 3C\,390.3 in 
1994--95 are presented. The broad-band fluxes ($B$, $V$, $R$, and $I$), 
the spectrophotometric optical continuum flux  $F_{\lambda}$(5177\,\AA), 
and the integrated emission-line fluxes of \Halpha, \Hbeta, \Hgamma, \HeI, 
and \HeII\ all show a nearly monotonic increase with episodes of milder 
short-term variations superposed. The amplitude of the continuum variations 
increases with decreasing wavelength (4400 --- 9000 \AA).
The optical continuum variations follow the variations in the ultraviolet 
and X-ray with time delays, measured from the centroids of the 
cross-correlation functions, typically around 5 days, but with
uncertainties also typically around 5 days; zero time delay between the
high-energy and low-energy continuum variations cannot be ruled out. The
strong optical emission lines \Halpha\,, \Hbeta\,, \Hgamma\,, and \HeI\
respond to the high-energy continuum variations with time delays typically
about 20 days, with uncertainties of about 8 days.
There is some evidence that \HeII\ responds somewhat more rapidly, with
a time delay of around 10 days, but again, the uncertainties are quite
large ($\sim $\,8 days).
The mean and rms spectra of the \Halpha\ and \Hbeta\ line profiles
provide indications for the existence of at least three distinct components
located at $\pm4000$ and 0\,\kms\ relative to the line peak.
The emission-line profile variations are largest near line center.
\end{abstract}
\section{Introduction}
Intensive multiwavelength monitoring campaigns have shown that variability
studies provide an excellent tool to investigate the innermost region of 
active galactic nuclei (AGNs). Detailed studies of the continuum 
and emission-line variations have revealed new insights about 
the size, structure, and dynamics of the broad-line region (BLR) 
in these sources (see Peterson 1993 for a review).

Over the last decade, a number of large space-based and ground-based AGN 
monitoring programs have been undertaken. Our own group, the International 
AGN Watch consortium (Alloin et al.\ 1994), has undertaken monitoring 
programs on several Seyfert galaxies, including programs on 
NGC 5548 (Clavel et al.\ 1991; Peterson et al.\ 1991, 1993, 1994;
Maoz et al.\ 1993; Dietrich et al.\ 1993; Korista et al.\ 1995),
NGC 3783 (Reichert et al.\ 1994; Stirpe et al.\ 1994; Alloin et al.\ 1995),
NGC 4151 (Crenshaw et al.\ 1996; Kaspi et al.\ 1996; Warwick et al.\ 1996;
Edelson et al.\ 1996), Fairall 9 (Rodr\'{\i}guez-Pascual et al.\ 1997; 
Santos-Lle\'{o} et al.\ 1997), and NGC 7469 (Wanders et al.\ 1997;
Collier et al.\ 1997).
In late 1994, the International AGN Watch began a multiwavelength 
monitoring campaign on the broad-line radio galaxy (BLRG) 3C\,390.3, 
a prominent nearby ($z = 0.0561$; Osterbrock et al.\ 1975) AGN with broad 
double-peaked emission-line profiles (Sandage 1966; Lynds 1968). 
3C\,390.3 has a well-known variability
history (e.g.\, Selmes, Tritton, \& Wordsworth 1975; Barr et al.\ 1980; 
Penston \& P\'erez 1984; Veilleux \& Zheng 1991; Zheng 1996; Wamsteker 
et al.\ 1997). Zheng (1996) and Wamsteker et al.\ (1997) analysed the
IUE spectra of 3C\,390.3 which have been taken from 1978 until 1992. The
variable broad \Lya\ and \civ\ emission are delayed by $\sim $60 days with
respect to the UV continuum variations.

3C\,390.3 is an extended double-lobed FR II radio source (Leahy \& Perley 
1995) with a 60\arcsec \,one-sided narrow jet at ${\rm PA} = -37^{\circ}$ 
and it is one of the rare lobe-dominated radio galaxies which show 
superluminal motions, with $v/c \approx 4.0$ (Alef et al.\ 1996).
This is the first time that variations in  a radio-loud AGN have been
studied over a broad energy range (from radio to X-ray energies) for
an extended period (one year) --- the observations obtained for this
program cover over eight decades in frequency. The results of the X-ray and 
ultraviolet monitoring campaigns performed with ROSAT and the IUE satellite 
will be presented in Leighly et al.\ (1997) and O'Brien et al.\ (1998), 
respectively.

In this contribution, we present the optical photometric and spectroscopic
observations that were obtained as part of this monitoring program;
observations in other wavelength bands will be presented elsewhere. In \S{2}, 
we describe the optical observations and outline intercalibration procedures 
by which a homogeneous set of the photometric and spectroscopic measurements 
is achieved. In \S{3}, we highlight emission-line profile variations based on 
mean and rms spectra and present the results of some preliminary time-series 
analysis. We summarize our results in \S{4}.

\section{Observations and Data Analysis}
Spectra and broad-band photometric measurements of 3C\,390.3 were obtained 
by a large number of observers between 1994 October and 1995 October.
Table 1 gives a brief overview of the various sources of the data we report 
here. Each group (column 1) was assigned an identification code given in 
column (2) which will be used throughout this paper. Column (3) gives the 
aperture of the telescope used. Columns (4) -- (7) list the focal-plane 
apertures used in various observations;
generally, fixed instrument apertures were used, although two groups (F, N) 
adjusted their aperture to compensate for changes in seeing. In most cases 
the size of the apertures corresponds to three times the average seeing at 
the observing site. Those large apertures cover the AGN as well as the entire 
galaxy. In column (8), the spectrograph slit widths (in the dispersion 
direction) and extraction widths (cross-dispersion dimension), respectively,
of the spectra are listed.

Complete logs of the photometric and spectroscopic observations (Table 2 
and 3) are available as electronic files on the WWW at the URL given in the 
following. 
http://www.astronomy.ohio-state.edu/$\sim $agnwatch/

\subsection{Optical Photometry}
The photometric observations were made through different combinations of 
broad-band filters. Generally, the brightness of 3C\,390.3 was scaled with 
respect to standard stars in the photometric sequence (stars A, B, and D) 
defined by Penston, Penston, \& Sandage (1971), plus a HST guide star located 
at 208\arcsec\ from 3C\,390.3 at ${\rm PA} = 106^{\circ}$ (HST GS 4591:731).
In some cases, star A could not be used because of saturation.

The colors of the comparison stars (B,D, and HST GS 4591:731) differ by up
to 0.1 mag.\, while star A is $\sim$2.5 mag.\, brighter than the mean values 
of these three stars. The mean colors of 3C\,390.3 are similiar to the mean 
colors of the comparison stars within 0.2 mag. Since the stars and the AGN 
are within the same field of view of the CCD images, effects of different 
spectral energy distributions 
and airmasses on the internal calibration can be neglected.

In the following, the procedure we use to derive the $R$-band magnitudes 
will be described in some detail. The light curves of the other broad-band 
measurements were obtained in a similar way.

\subsubsection{$R$-band:}
First, a subsample was selected which covered the entire monitoring campaign 
with a good temporal sampling rate. For the $R$-band measurements, the 
observations recorded with the 0.7-m telescope of the Landessternwarte 
Heidelberg (L) provide an appropriate data set. The used aperture was a 
15\arcsec\ rectangle to ensure that the light loss can be neglected even in 
the case of bad seeing. The frames were bias and flat-field corrected in the 
standard way. In the next step, the brightness of 3C\,390.3 was measured with 
respect to the comparison stars in the same field, yielding the difference in 
apparent magnitude between 3C\,390.3 and the comparison stars. In the 
$R$-band, stars B and D from Penston et al. (1971) and HST GS 4591:731 
mentioned above were used for calibration. The $R$-band magnitudes that were 
derived for these stars are presented in Table 4 together with the $B$ and 
$V$ magnitudes from Penston et al.\,(1971). Finally, the $R$-band light 
curves of the other  subsamples were shifted to the $R$-band light curve 
derived from the Heidelberg sample by applying a constant additive magnitude 
offset to all of the measurements in a given subset. The additive factor was 
derived by comparison of $R$-band magnitudes which were observed within 
$\pm$3 days. The additive factors for the individual subsamples are 
presented in Table 5.
The additive scaling factors given in Table 5 are quite large in some cases.
Generally, the brightness variations of 3C\,390.3 were provided in magnitudes
using the standard stars in the field for calibration. But a few groups 
(M, R, U) provided the brightness of 3C\,390.3 as differential magnitudes
relative to the calibration stars in the field.

\subsubsection{$B$-band:}
The $B$-band magnitudes are scaled with respect to the measurements taken at 
Wise Observatory (O). The offset for the Steward Observatory sample (F) was 
derived by comparison of epochs which were simultaneous to  within $\pm$3 
days. This restriction could not be used for the Special Astrophysical 
Observatory (J); in this case, the offset is based on epochs separated by no 
more than 6 days. The resulting additive scaling factors are given in Table 5.
\subsubsection{$V$-band:}
The sample recorded at Behlen Observatory (Q) was used as the standard in
the $V$-band because the temporal sampling is good and the measurements
were made through a large aperture (13\arcsecpoint7)
so that uncertainties due to seeing variations are small. The $V$-band
measurements are made relative to star B, which is close in brightness
to the nucleus of 3C 390.3.
In order to scale  the $V$-magnitudes to a common flux level, 
we compared measurements at epochs separated by no more than three days.
In Table 5, the differential magnitudes are given which were applied to the 
individual light curves to produce a common light curve.
\subsubsection{$I$-band:}
For the calibration of the $I$-band magnitudes we used observations taken 
at Calar Alto Observatory. The $I$-band frames were
recorded in December 1994. The calibration was based on the globular cluster
NGC\,2419 using the stars given by
Christian et al.\,(1985). The resulting brightnesses of star A, B, D, and
HST GS 4591:731 are given in Table 4. The additive factors for the 
individual subsamples are presented in Table 5.

The measured brightnesses of the stars used for calibration are in good 
agreement with the values provided by Penston et al. (1971).
However, star A deviates from this trend in $R$ and $I$ in comparison
to stars B and D. It might be that this is due to the brightness of star A 
causing non-linearity effects and an underestimate of the brightness of star
A. After combining the measurements of the individual groups to common 
broad-band light curves, the apparent magnitudes were transformed into flux. 
The conversion has been performed using the following equations (Allen 1973; 
Wamsteker 1981):
\begin{eqnarray}
\log F_\lambda (4400\,{\rm \AA}) & = & -0.4 m_B - 8.180,   \\
\log F_\lambda (5500\,{\rm \AA}) & = & -0.4 m_V - 8.439, \nonumber \\
\log F_\lambda (7000\,{\rm \AA}) & = & -0.4 m_R - 8.759, \nonumber \\
\log F_\lambda (9000\,{\rm \AA}) & = & -0.4 m_I - 9.080. \nonumber
\end{eqnarray}
The resulting time-binned light curves from the $B$, $V$, $R$, and $I$ 
broad-band measurements in 
units of $10^{-15}$\,ergs\ s$^{-1}$\,cm$^{-2}$\,\AA$^{-1}$ are 
displayed in Figure 1.
\subsection{Optical Spectroscopy}
The flux calibration of AGN spectra can be accomplished in several ways.
In variability studies, it is common practice to normalize the flux scale 
to the fluxes of the narrow emission lines, which are assumed to be constant 
over time scales of at least several decades (Peterson 1993). This assumption 
is justified by the large spatial extent and low gas density of the
narrow-line region (NLR), since light travel-time effects and the long
recombination time scale ($\tau_{\rm rec} \approx 100$\,years
for $n_e \approx 10^3$\,cm$^{-3}$) damp out
short time-scale variability. However, 3C\,390.3 is apparently a
special case in this regard, as narrow-line variability has been
reported in this object (Clavel \& Wamsteker 1987). 
On the basis of spectra obtained between 1974 and 1990,
Zheng et al.\ (1995) present evidence that 
the [\oiii]\,$\lambda\lambda4959$, 5007 fluxes follow the variations
in the continuum, although on a longer time scale than the 
broad emission lines. 
During the period of decreasing and increasing [\oiii]\ flux there
might also be periods of several months or one year of nearly constant  
[\oiii]\ flux.
However, we need to examine the data closely to test for variability of the 
[\oiii]\ lines before we use them for flux calibration.

Two data sets, the Ohio State sample (A) and the Lick Observatory 
sample (B), were selected to measure the absolute flux in the
\ob\ line. The spectra were photometrically calibrated 
by comparison with the broad-band photometric 
measurements. The spectra were convolved with the spectral response curve of 
the Johnson $V$ and $R$ filters (Schild 1983), and  the flux of the
convolved spectra was measured for the wavelength range of the filter 
curves and compared with photometric data points. Photometric and
spectroscopic data obtained no more than 3 days apart were used in this
comparison.
The spectra were then scaled by multiplicative factors to 
achieve the same total flux ratio as the intercalibrated photometric 
measurements in $V$ and $R$. For the Lick data (B),
the $R$-band measurements were used to 
scale the entire spectrum, and the \ob\ flux was then measured.
Spectra from the Ohio State sample (A) do not cover
the entire wavelength range of the $V$-band filter, and therefore the
missing contribution to the $V$-band measurement was estimated from
the Lick spectra, which cover the entire wavelength range of the
$V$ bandpass. The ratio of $V$-band flux measured from
the Ohio State spectra to that measured from the Lick spectra
was found to be $0.830\pm0.002$,  i.e., a
constant contribution. The fluxes from the Ohio State spectra were 
thus corrected by this constant factor.

The \ob\ flux was then measured from the photometrically
scaled Ohio State (A)
and Lick Observatory (B) spectra by integrating the spectrum over
the range 5258--5320\,\AA. 
Therefore, a linear pseudo-continuum was fitted beneath the \ob\
emission line. In Fig.\ 2, we show the
\ob\ flux measured from the Ohio State and Lick subsets
normalized to a mean value of unity and displayed as a function of time. 
No time-dependent trend is detected, and the rms variation about the mean 
is $\sim2.6$\%. We thus conclude that it is safe to assume that the 
\ob\ flux is constant within a few percent over the duration of this 
monitoring program, and that the \ob\ flux can be used to calibrate all of 
the spectra. The \ob\ flux is taken to be
$F(\mbox{\rm [O\,{\sc iii}}]\,\lambda5007) = 
1.44 \times 10^{-13}$\,ergs\ s$^{-1}$\,cm$^{-2}$,
the mean value of the data points plotted in Fig.\ 2.

It is also important to take aperture effects
into account (Peterson \& Collins 1983). The 
seeing-dependent uncertainties which are
introduced by the aperture geometry can be minimized
by using large apertures. It has been shown that apertures
of 5\arcsec\ $\times$ 7\arcsecpoint5 
can reduce seeing-dependent photometric errors to no more than a few percent 
in the case of nearby AGNs (Peterson et al.\ 1995).
In the case of 3C\,390.3, the BLR and the NLR can be taken to be  point 
sources (cf.\ Baum et al.\ 1988),
which means that no aperture correction needs to be made for the
AGN continuum/narrow-line flux ratios or the broad-line/narrow-line ratios,
since seeing-dependent light losses at the slit will be the same for each
of these components. However, the amount of host-galaxy starlight that
is recorded is still aperture dependent, and  systematic corrections need
to be employed.

In order to estimate how sensitive the measurements are to
seeing-dependent light loss from the host galaxy,
we carried out simulated aperture photometry on the $R$-band frame
obtained at Calar Alto with the 2.2-m telescope under good seeing 
conditions ($\sim0\arcsecpoint7$). The host galaxy has been separated from the
point-like AGN component by fitting a
de Vaucouleurs $r^{1/4}$ profile to the observed surface-brightness
distribution. This image was convolved with Gaussians of various
width to simulate various seeing conditions up to 
4\arcsecpoint0. 
The flux of the host galaxy and of the point-like AGN were measured for
a fixed aperture of 10\arcsecpoint5\,x\,10\arcsecpoint5.
The flux ratio of the point-line AGN to the host galaxy for the $R$-band is
$F_{\rm AGN}/F_{galaxy} = 0.63 \pm 0.03$. This is similar to the result of 
Smith \& Heckman (1989), who found $F_{\rm AGN}/F_{galaxy} = 0.44$ for 
$V$-band measurements.
By using different aperture geometries in these simulations,
we find that the ratio of AGN light to starlight from the host galaxy
changes over the full range of observed seeing values by less than 1\% for
the larger apertures (i.e., slit width greater than 4\arcsec).
For intermediate slit widths (2\arcsecpoint5 
-- 3\arcsecpoint6), seeing variations
introduce uncertainties of $\sim5$\%. 
For the smallest slit widths ($\leq 2\arcsecpoint5$), seeing effects
can alter the nucleus to starlight ratio by as much as 
$\sim10$\%, with the largest uncertainties occurring 
for seeing worse than $\sim3\arcsec$.
\subsection{Intercalibration of the Spectra}
Since the data that constitute the various samples were taken with different
instruments in different configurations, 
the spectra have to be intercalibrated to a common
flux level. As we have shown above, the \ob\ line flux was constant 
to better than 3\% during this campaign, so we can safely use
the narrow emission lines as flux standards. In order to avoid any
wavelength-dependent calibration errors, each spectrum was scaled in flux
locally over a limited wavelength range prior to measurement.
The \Hbeta\ spectral region was scaled with respect to the 
[\oiii]\,$\lambda\lambda4959$, 5007 line fluxes, 
while the \Halpha\ region was scaled with respect to the fluxes of the
[\oi]\,$\lambda6300$ and [\nii]\,$\lambda\lambda6548$, 6584 emission lines. 
The spectra were intercalibrated using the method described by van 
Groningen \& Wanders (1992). This procedure corrects the data for different 
flux scales, small wavelength shifts, and different spectral resolutions
by minimizing the narrow-line residuals in difference spectra formed by
subtracting a ``reference spectrum'' from each of the observed spectra.
The rescaled spectra are used to derive integrated emission-line fluxes 
as well as the optical continuum flux.

The continuum fluxes are then adjusted for different amounts of host-galaxy 
contamination (see Peterson et al.\ 1995 for a detailed discussion) through 
the relationship
\begin{equation}
F_{\lambda}(5177\,{\rm \AA}) = F_{5007} 
\left[ \frac{F_{\lambda}(5177\,{\rm \AA})}
{F(\mbox{\rm [O\,{\sc iii}]}\,\lambda5007)} \right]_{\rm obs} - G,
\end{equation}
where $F_{5007}$ is the adopted absolute \ob\ flux,
the quantity in brackets is the observed continuum to
\ob\ flux ratio measured from the spectrum, and
$G$ is an aperture-dependent correction for the host-galaxy flux.
The Ohio State sample (A), which uses a relatively large aperture
(5\arcsec $\times$ 7\arcsecpoint5), was adopted as a standard
(i.e., $G = 0$ by definition), and other data sets were merged
progressively by comparing measurements based on observations made no more than
$\pm$\,3 days apart. This means that any real variability that occurs on time 
scales this short tends to be somewhat suppressed by the process that allows 
us to combine the different data sets. 
The additive scaling factor $G$ for the various samples are given in Table 6.
\section{Results}
\subsection{Light Curves}
The average interval between measurements is about $2\pm0.5$\,days for 
$V,$ $R,$ and $I$ while the $B$-band variations have been measured with an 
average sampling interval of $16\pm18$\,days (Table 7).
The time-binned broad-band continuum light curves are shown in Fig.\ 1. 
The variations can be characterized as a nearly monotonic 
increase of the flux with smaller-scale variations superposed on this
general trend.  From JD2449760 to JD2449800, the flux rose in the V,\,R,\,I 
bands faster than during the previous interval. 
After this period, the flux 
level stays nearly constant for nearly 3 months and then a second strong 
increasing episode follows. The $V$-band light curve appears to have
more complicated structure, and we note in particular apparently
rapid variations during the interval JD2449800 to JD2449920.

Figure 3 shows two spectra of 3C\,390.3 that represent the low (JD2449636) 
and high (JD2450007) flux states observed during this campaign. 
The most obvious variation is the strong increase of the small blue bump,
the very broad feature shortward of 4000\,\AA, 
which is usually ascribed to Balmer continuum emission and a blend of several
thousand \feii\ emission lines (Wills, Netzer, \& Wills 1985). The
broad Balmer lines show also evidence for increasing flux.
Underneath the low-state spectrum in Fig.\ 3 we show the integration
ranges for the various features that have been measured in these spectra.
The optical emission-line fluxes were integrated over a common 
range of $\pm7500$\,\kms\ for the strong Balmer emission lines 
(\Halpha, \Hbeta, and \Hgamma) and \HeI. The \HeII\ line flux
was integrated over a range corresponding $-7500$\,\kms\ to $+4000$\,\kms\ 
to reduce contamination by \Hbeta \,(cf.\,Table 8). A local linear continuum 
fit was interpolated under each emission line. In the case of the \Hbeta\ 
region, the continuum 
was defined by the flux measured in two narrow (10\,\AA\ width) 
windows at 4400\,\AA\ and 5475\,\AA\ in the observed frame.
Although the region of the long wavelength window might be 
contaminated by several weak emission lines and the Mg\,b absorption feature,
this is still the most line-free region near \Hbeta\ and is thus
an appropriate place to estimate the optical continuum flux.
The continuum window at 4400\,\AA\ is displaced from the \Hgamma\ line
center by $\sim 12000$\,\kms, and thus possible contamination by broad 
emission line flux can be safely ignored (see Fig.\ 3). Contamination of 
\Hbeta\ by [\oiii] emission has been corrected by subtracting
the constant \ob\ flux given earlier 
plus the [\oiii]\,$\lambda4959$ flux which we account for by
assuming a \ob/[O\,{\sc iii}]\,$\lambda4959$ flux ratio of 3.
In the case of the \Halpha\ region, two narrow (10\,\AA\ in width) 
windows at 5960\,\AA\ and 7495\,\AA\ were used to define the continuum 
underlying the lines. No attempt has been made to correct any of the 
measured emission-line fluxes for their respective narrow-line contributions. 
In each spectrum, the optical continuum flux measured is the average value 
in the range 5460--5470\,\AA. The extracted range of the emission lines and 
of the optical continuum is given in Table 8.

The time-binned light curves for the strong emission lines and the optical 
continuum flux $F_{\lambda}$(5177\,\AA) measured from the calibrated spectra 
are plotted in Fig.\ 4. The corresponding flux measurements are available
on the WWW at the URL given above (Table 9).
 
A final check of the uncertainty estimates was performed by examining the 
ratios of all pairs of photometric and spectroscopic observations which were 
separated by 2 days or less. The error estimates were also calculated using 
maximum intervals of 3 days and 4 days, and the resulting errors are 
identical to within 3\%.
Generally, there are at least a few hundred independent pairs of measurements 
within 2 days of one another for the photometric measurements. Only the 
$B$-band curve yields fewer independent pairs. 
For the spectroscopic data the number of independent pairs is of the order
of several dozen with the exception of \Hgamma .
The dispersion about the mean (unity), divided by $\sqrt 2$, provides an 
estimate of the typical uncertainty in a single measurement 
($\sigma _{\rm est}$). 
The observational uncertainties ($\sigma _{\rm obs}$) assigned to 
the spectral flux measurements were estimated from the error spectra which 
were calculated within the intercalibration routine, as well as from the 
signal-to-noise ratio within the spectral range near the individual emission 
lines. For the continuum the mean fractional error ($\sigma _{\rm obs}$) in 
a given measurement is 0.041. The average fractional uncertainty from the 
quoted estimate ($\sigma _{\rm est}$) for the same measurements (Table 10) 
is 0.040 which implies that the error estimates are probably quite good. 
Generally, the estimated errors ($\sigma_{\rm est}$) are of the same order
as the observational uncertainties ($\sigma_{\rm obs}$) derived directly 
from the measurements.

A comparison of the flux of the broad band measurements and
the broad emission lines is given in Table 11.
The mean spectroscopic continuum flux is lower than the mean broad band
$V$- and $R$-flux. This can be explained by the fact that the broad band
flux measurements contain in addition to the continuum flux emission line
contributions.
The variability parameter $F_{\rm var}$ and $R_{\rm max}$ have been
calculated for the broad-band flux variations (cf.\ Clavel et al.\ 1991;
Rodr\'{\i}guez-Pascual et al.\ 1997). The quantity $R_{\rm max}$ is simply the 
ratio of the maximum to the minimum flux. The quantity $F_{\rm var}$ is an
estimation of the fluctuations of the intrinsic variations relative to
the mean flux. Therefore, the rms of the light curves has been corrected
with respect to the uncertainties introduced by the observations. 
$R_{\rm max}$ and $F_{\rm var}$ of the broad-band variations are given in 
Table 11.
\subsection{Mean and Root-Mean-Square Spectra}
We calculated the mean and root-mean-square (rms)
spectra from the flux-scaled spectra,
and the \Halpha\ and \Hbeta\ regions are shown in Fig.\ 5. The velocity 
scale is set by adopting a redshift $z = 0.0556$ given by the narrow 
components of \Halpha\ and \Hbeta\ and the [\oiii]\,$\lambda4959,5007$
emission lines, i.e.\ the restframe of the NLR. The rms spectrum is useful 
for isolating the variable parts of the line profile. 
The full-width at zero intensity (FWZI) of the mean \Halpha\ and \Hbeta\
profiles is (25000$\pm$2000)\,\kms, while the FWZI of the rms profiles is 
only (14000$\pm$500)\,\kms\, (cf.\,Fig.\,5). Thus, the variations in the broad 
emission line profiles are strong at lower radial velocities while the flux 
originating in high radial-velocity gas varies little, if at all. This 
behavior is similar to what has been seen in Mrk 590 
(Ferland, Korista, \& Peterson 1990; Peterson et al.\ 1993).

The line profiles of \Halpha\ and \Hbeta\  show clear asymmetric structure.
At least three substructures can be identified in both 
line profiles. A blue hump in the mean spectra is located near 
$-3500$\,\kms\ with respect to the line peak, 
and a red hump is seen near $+4300$\,\kms.
The red and blue humps can be seen clearly in the rms spectra,
and in addition a broad central component appears. 
If these three components are modeled as Gaussians, the
FWHM of the components in the wings is of order $3000$\,\kms,
while for the central component the FWHM is $\sim4000$\,\kms.
The components in the profile wings are nearly symmetrically 
located with respect to the line center. 

In the rms-spectrum of the \Hbeta\ emission line a strong broad feature 
at $\sim $\,10000\,\kms\, is clearly visible. The FWHM of the structure
can be estimated to $\sim $\,3500\,\kms. The residuals of the narrow
[\oiii] lines are located at $\sim $\,6000\,\kms\,\ and $\sim $\,8900\,\kms. 
Since this feature is also visible in the rms spectrum of \Halpha\ and
of \Hgamma\ it cannot be caused entirely by inaccurate scaling of the 
NLR contribution of the narrow [\oiii] lines. Furthermore, in the difference
spectrum of the low and high state of 3C\,390.3 during this monitoring 
campaign (cf.\ Fig. 3) a strong broad feature is clearly visible at the
red side of \Halpha\ ($\sim $\,7150 \AA ) and a weaker structure at the red 
side of \Hbeta\ ($\sim $\, 5300 \AA ). The analysis of the simultaneous
ultraviolet campaign reveals the existence of a feature at 
$\sim $\,8500\,\kms\,\ in the outer red wing of the \civ\,$\lambda $1548 
emission line (O'Brien et al.\,1998).
\subsection{Time-Series Analysis}
In order to quantify any possible time delay between the various light 
curves shown in Figs.\ 1 and 4, we perform a simple cross-correlation 
analysis. Three methods that are commonly used in AGN variability studies
have been used to compute cross-correlation functions (CCFs) --- 
the interpolated cross-correlation function (ICCF) of Gaskell \& Sparke (1986)
and Gaskell \& Peterson (1987),
the discrete correlation function (DCF) of Edelson \& Krolik (1988),
and the $z$-transformed discrete correlation function method (ZDCF) of
Alexander (1997). The ICCF and DCF algorithms and the limitations of the
methods have been discussed in detail by Robinson \& P\'erez (1990) and 
by White \& Peterson (1994), and the specific implementation of the
ICCF and DCF used here are as described by White \& Peterson (1994).

The emission lines are expected to change in response to variations in
the far-UV continuum, primarily to the unobservable wavelengths just 
shortward of 912\,\AA. We must therefore assume that the observable 
continuum can approximate the behavior of the ionizing continuum. It 
is generally assumed in this type of analysis that the shortest observed UV 
wavelength provides the best observable approximation to 
the ``driving'' (ionizing) continuum. For the first time,
however, we have a well-sampled simultaneous soft X-ray light curve
(Leighly et al.\ 1997), so we can compare the X-ray and UV continua
(O'Brien et al.\ 1998) directly. In this analysis, we used the X-ray 
light curve as the driving continuum (Fig.\,6) since the observational 
uncertainties are smaller than for the UV light curve. As will be discussed 
elsewhere (O'Brien et al.\ 1998), the relative amplitudes of the
ultraviolet variations at a wavelength of 1370\,\AA\ are
nearly identical to those in the {\em ROSAT}\/ HRI light curve,
so the results presented here do not depend critically on our choice
of the X-ray light curve in preference to the UV light curve.

Since the sampling of the $B$-band light curve is very poor,
we have excluded the $B$-band data from the cross-correlation analysis.
Operationally, the
temporal coverage of the optical light curves was restricted to the
period covered by the X-ray measurements. All of the light curves
were binned into time intervals of 0.5 days to avoid unnecessary
structure in the ICCF. For each light curve, we also computed
the sampling-window autocorrelation function (ACF$_{\rm sw}$),
which is a measure 
of how much of the width of the ACF is introduced by the interpolation 
process rather than by real correlation of the continuum values at different
times. The ACF$_{\rm sw}$ is computed by repeatedly
sampling white-noise light curves in exactly the same way as the real 
observations and then computing the autocorrelation function (ACF).
The average of many such autocorrelations, the ACF$_{\rm sw}$,
has a peak at zero lag whose width depends on how much interpolation
the ICCF has to do on short time scales.
The width (FWHM) of the ACF$_{\rm sw}$ is $1.8\pm0.3$ days
for the emission lines and broad-band flux measurements for the
X-ray restricted time period. These values are 
negligible compared to the widths of any of the
emission-line ACFs or CCFs found here (see Figs.\ 7--9), and thus
interpolation of the light curves is justified.

The general sampling characteristics of each of the restricted and rebinned
light curves are given 
in Table 12. The name of the feature is given in column (1), and column (2)
gives the total number of points $N$ in the light curve
that are used in computing the cross-correlation functions. The width 
(FWHM) of the ACF is given in column (3), and the width 
of ICCF computed by cross-correlation with the X-ray continuum is given in 
column (4). Column (5) gives the width (FWHM) of the corresponding 
ACF$_{\rm sw}$.

Uncertainties in the ICCF results for the cross-correlation maxima
and centroids, $\Delta\tau_{\rm max}$ and $\Delta\tau_{\rm cent}$,
respectively, were computed through Monte Carlo techniques as follows:
for both time series, each flux value in the light curves was modified with 
Gaussian deviates based on the quoted uncertainty for that point.
Each light curve of $N$ points was then randomly resampled $N$ times
in a ``bootstrap'' fashion, specifically allowing points to appear more
than one time (the effect of which is only to remove at random
certain points from the light curve). The ICCF was computed, and
the values of $\tau_{\rm max}$ and $\tau_{\rm cent}$ were recorded
if they were statistically significant at a confidence level higher
than 95\% and not clear outliers (e.g., lags larger than 100 days).
By repeating this process 500--1000 times, distributions of
$\tau_{\rm max}$ and $\tau_{\rm cent}$ were built up. The means of
these distributions were always very close to the values
of $\tau_{\rm max}$ and $\tau_{\rm cent}$ obtained from the original
series, and the standard deviations of these distributions are
taken to be the uncertainty associated with a single realization,
i.e., the quoted uncertainties
$\Delta \tau_{\rm max}$ and $\Delta\tau_{\rm cent}$.

The ACF, ACF$_{\rm sw}$, ICCF, and ZDCF are shown  for the broad-band 
variations in $V$, $R$, and $I$ (Fig.\ 7), 
the broad Balmer emission lines \Halpha, \Hbeta, and \Hgamma\ (Fig.\ 8), 
the helium lines \HeI\ and \HeII, and the optical continuum 
$F_{\lambda}$(5177\,\AA) (Fig.\ 9).
The ACFs of the light curves are broad since the shape is dominated by the 
nearly monotonic increase in the light curves. The FWHMs of the ACFs of the 
well-sampled light curves are of the order of 50--100\,days (Table 12). 
Within the uncertainties, the DCFs and ZDCFs are identical; therefore,
to avoid confusion only the ICCFs and the corresponding ZDCFs are 
displayed in Figs.\ 7--9.

The time delay derived from the centroid of a cross-correlation function
provides a more robust estimate of the lag than does the peak, as evidenced 
by the consistently smaller widths of the Monte Carlo distributions for the
centroid compared to the peak. Also, in the case of the emission lines,
the centroid is readily identified with a physically meaningful quantity, the
luminosity-weighted radius of the line-emitting region 
(Koratkar \& Gaskell 1991).
The delays expressed by $\tau _{\rm cent}$ of the three methods used here are
nearly identical within the uncertainties, although the DCF method tends to 
yield smaller delays than the ICCF and ZDCF method.

The results of the cross-correlation analysis are given in Table 13.
Column (1) indicates the ``responding'' light curve (i.e., the light
curve that is assumed to be responding to the driving light curve), and 
column (2) gives the peak value of the correlation coefficient $r_{\rm max}$ 
for the ICCF. The position of the peak of the cross-correlation functions 
$\tau _{\rm max}$ was measured by fitting a Gaussian curve to the upper 
85\% of the ICCF, ZDCF, and DCF; these values are given in columns (3), (4), 
and (5), respectively, and column (6) gives the error estimate for
the position of the cross-correlation peak $\Delta \tau_{\rm max}$. 
The centroids $\tau _{\rm cent}$ of the ICCF, ZDCF, and DCF,
in each case computed using the points in the cross-correlation
function with values greater than $0.8r_{\rm max}$, are given in
columns (7), (8), and (9), respectively, and the uncertainty in
the ICCF centroid $\Delta \tau_{\rm cent}$ is given in column (10).

The broad-band ($V$, $R$, and $I$) 
and the optical continuum flux variations appear to be delayed by 
a few days, relative to the X-ray or ultraviolet continuum
variations. However, the measured lags are in no case
different from zero at any reasonable level of 
statistical significance if the Monte-Carlo based error
estimates are reliable. Somewhat smaller, 
but marginally statistically significant, wavelength-dependent continuum
lags have been reported in the case of NGC 7469
(Wanders et al.\ 1997; Collier et al.\ 1997), although in the
case of NGC 7469 the mean spacing between observations of the
driving continuum is much smaller than for the observations reported here.

The emission-line time delays are similar for all of the 
optical emission lines, about 20 days, although in each case the 
uncertainties are somewhat larger than we have obtained in
similar experiments on account of the vagaries of the
continuum behavior and sampling. Within the uncertainties,
the measured time delays for \Halpha, \Hbeta, \Hgamma,
and \HeI\ are indistinguishable. However, the uncertainties
are sufficiently large that ionization stratification, as
detected in other well-studied sources, also cannot be ruled out.
Only \HeII\ appears to respond more rapidly than the other lines,
but the relatively low value of $r_{\rm max}$ and relatively
narrow width of the cross-correlation function (about 30 days as compared
to $\sim 50$ days for the other lines) cast some doubt on
the significance of this result. In contrast to the AGN studied before,
the delay of the \Lya\, and \civ\, line variations of the current 3C\,390.3
campaign is significantly larger than the delay of the optical lines.
The analysis of the ultraviolet spectra yields a delay of 35 --- 70 days
for \Lya\,\ and \civ\,\ (O'Brien et al.\ 1998).

In order to study the response of individual parts of the line
profile to the X-ray continuum variations and thus search for
evidence of an organized radial-velocity field in the BLR,
we have divided the profiles of the strongest and least contaminated
lines, \Halpha\ and \Hbeta, into three parts: 
the blue wing ($-7500$ to $-2000\,$km\,s$^{-1}$),
the core ($-2000$ to $+2000\,$km\,s$^{-1}$), and 
the red wing ($+2000$ to $+7500\,$km\,s$^{-1}$).
The light curves of the different profile
sections (cf.\,Figs.\ 10,11) were rebinned onto 0.5-day intervals 
(available in electronic form on the WWW at the URL given above as
Table 14) and again
cross-correlations were performed restricting the data to
the X-ray monitoring period. Since we are concerned at this
point with {\it differential}\/ lags between different sections of
the line profiles, we cross-correlate the light curves for
the different line sections against each other rather than
against the driving continuum; we arbitrarily chose the line
cores as the first (driving) series, and computed the lags of the red and
blue  wings of the lines relative to the core.
These results are shown in Table 15 and in Figs.\ 12 and 13. 
Neither \Halpha\ nor \Hbeta\ shows any evidence for any
differences in the response of the wings relative to the core.
In the case of \Halpha, the uncertainties are especially large on
account of the low amplitude of variability, but the
\Hbeta\ results seem to exclude the possibility that
the BLR velocity field is characterized by primarily
radial motions, either infall or outflow.
\section{Summary}
The results of a year-long (1994 October to 1995 October)
optical monitoring campaign on the BLRG 3C\,390.3 are
presented in this paper. The principal findings are as follows:
\begin{enumerate}
\item The broad-band ($B$, $V$, $R$, and $I$) fluxes,
the optical continuum measured from spectrophotometry $F_{\lambda}$(5177\,\AA),
and the integrated emission-line fluxes of \Halpha, \Hbeta, \Hgamma,
\HeI, and \HeII\ showed significant variations of order
50\% in amplitude.
\item The parameter $F_{\rm var}$, which is essentially the rms
variation about the mean, increased with decreasing wavelength for the
broad-band measurements as well as for the Balmer emission lines.
\item The variations of the broad-band and emission-line fluxes 
are delayed with respect to the X-ray variations. Cross-correlation
functions were calculated with three different methods (ICCF, DCF, 
and ZDCF). 
The time delays of the optical continuum variations expressed by the
centroid of the cross-correlation functions are typically about 5 days,
but with uncertainties of $\sim $\,5 days. Therefore, zero-time delay
between the high-energy and low-energy continuum variability cannot be
ruled out.
The delays of the Balmer lines \Halpha, \Hbeta, and \Hgamma\, and of \HeI\
are typically around 20 days $\pm$ 8 days. There is some evidence that
\HeII\ responds somewhat more rapidly with a time delay of $\sim $\,10 days,
but again the uncertainties are quite large ($\sim $\,8 days).
\item The simple cross-correlation analysis of the line core with the
line profile wings of the \Halpha\, and \Hbeta\, emission might indicate that
the wings vary in the same way with respect to the line core. But there
might be a weak indication that the variations of the blue wing are delayed 
by $\sim $ 4 days with respect to the red wing.
\item The mean and rms \Halpha\ and \Hbeta\ line profiles
reveal the existence of at least three substructures --- a central
component, plus strong blue and red components
at about $\pm4000$\,\kms\ relative to line center.
The rms spectra show that the broad-line
variations are much stronger at line center than in the outer wings.
\end{enumerate}
\acknowledgments{ 
This work has been supported by
SFB328D (Landessternwarte Heidelberg), by the 
NSF through grant AST-9420080 (Ohio State University), 
                  AST 93-20715 (University of Arizona),
                  AST-9417213 (University of California, Berkeley),
          DFG grant KO 857/18-1 (Universit\"{a}tssternwarte G\"{o}ttingen),
          GSU Research Program Enhancement NASA grant NAGW-4397,
          Royal Society (University of Hertfordshire),
   Chinese NSF and Open Laboratory for Optical Astronomy of Chinese Academy
    of Sciences (Beijing Astronomical Observatory), 
   PPARC research studentship and PPARC grant GRK/46026 (St.\,Andrews),
 Russian Basic Research Foundation grant N 94-02-4885a, N 97-02-17625
 (Sternberg Astronomical Institute, Special Astrophysical Observatory),
 ESO C\&EE grant A-01-057,
 National Science Foundation Research Experience for Undergraduates (REU): 
 grant AST 94-23922, 
 and by the
 Valdosta State University Center for Faculty Development and Instructional
 Improvement grant. 
 We thank the FLWO observers P.\,Berlind and J.\,Peters (J.\,Huchra).}

\newpage

\figcaption[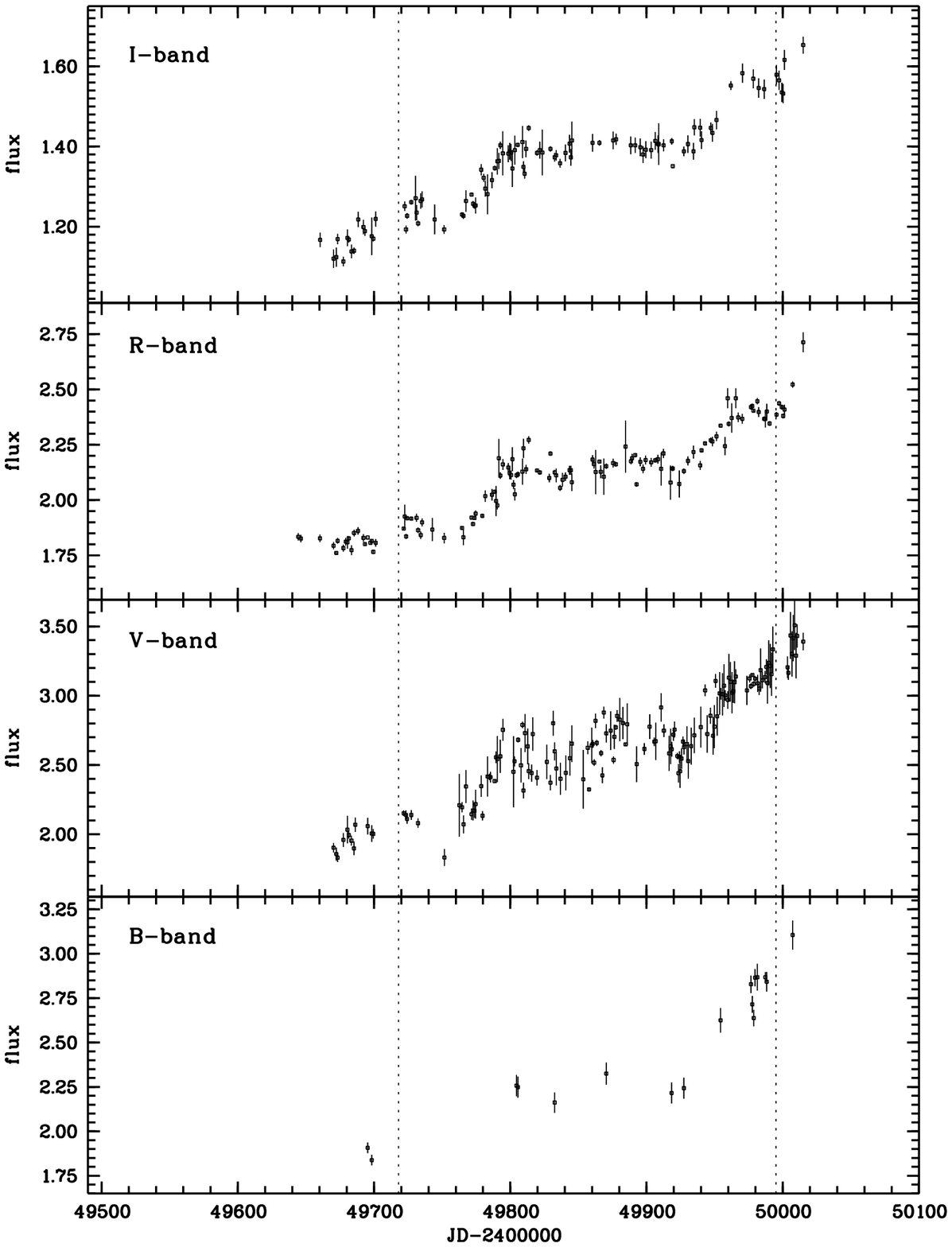]{Optical broad-band light curves of 3C\,390.3 from 1994 
October to 1995 October. Fluxes are in units of 
$10^{-15}$\,ergs\ s$^{-1}$\,cm$^{-2}$\,\AA$^{-1}$.
The larger uncertainties of the V-band measurements might be due to a lower
signal than in the R-band measurements which contains the broad 
\Halpha\, emission. 
The dashed vertical lines mark the temporal range of the X-ray observations.}

\figcaption[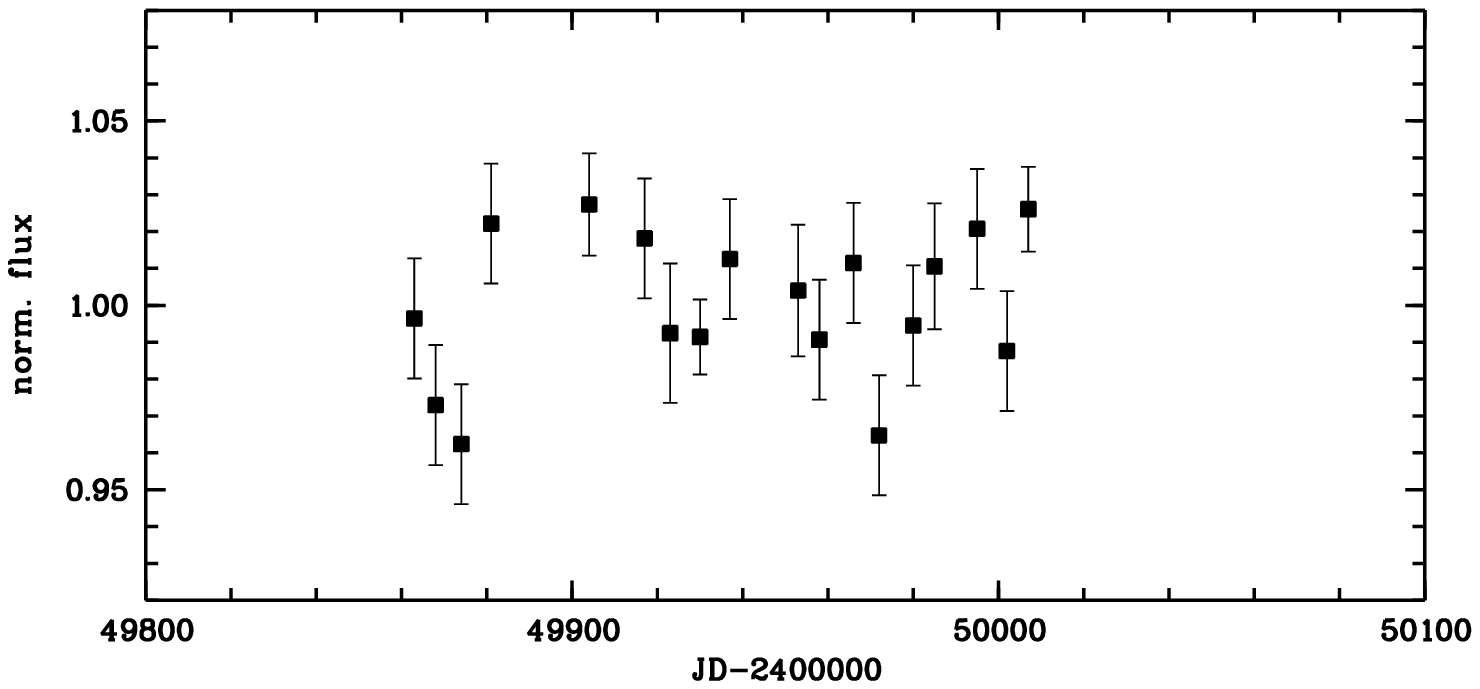]{Normalized \ob\ flux measured from photometrically
calibrated spectra from Ohio State (A) and Lick Observatory (B)
shown as a function of time. The data show no clear time dependence,
and the scatter about the mean is about 2.6\%. The narrow-line
[\oiii] flux thus appears to be constant over the time scales of
interest, and can be used to calibrate the fluxes of all the spectra.}

\figcaption[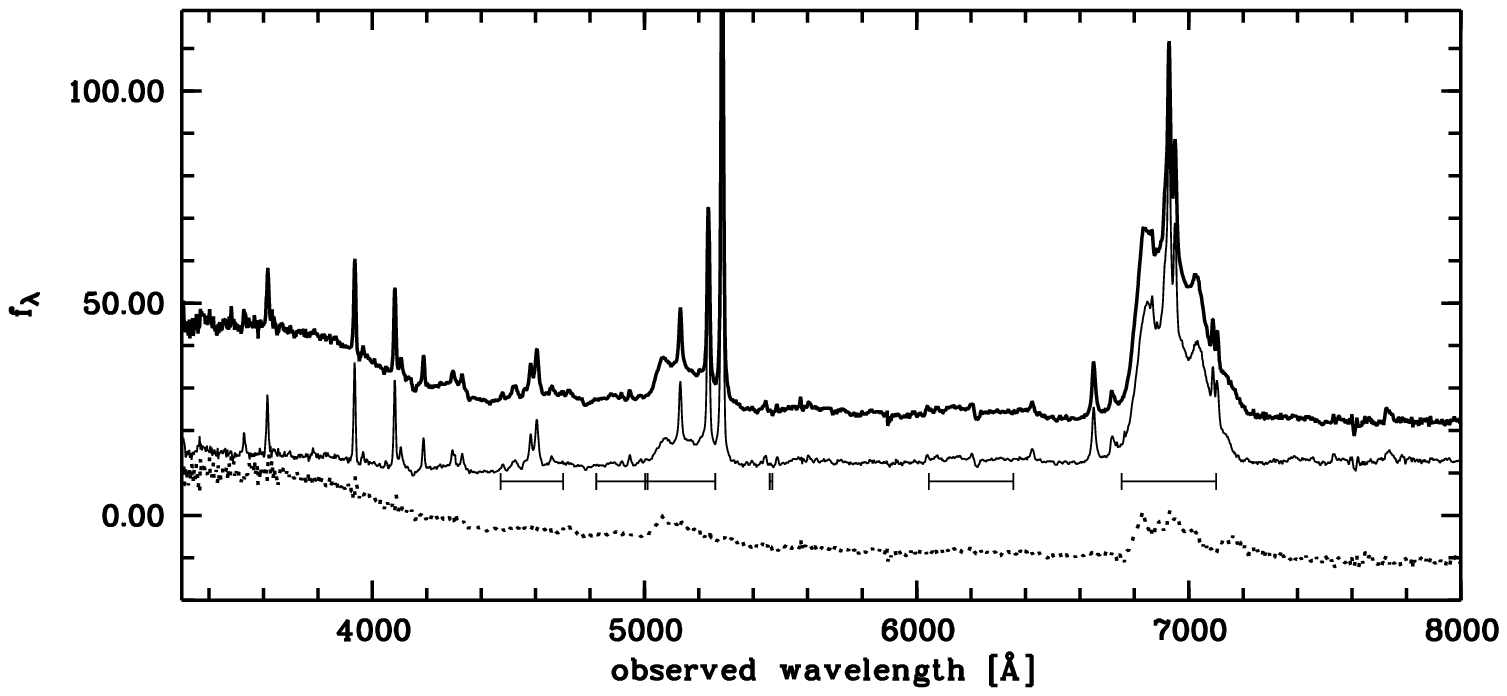]{Spectra from Lick Observatory (B) that illustrate the
high-flux (JD2450007, thick line) and low-flux (JD2449636, thin line) 
extremes observed during this campaign. The integration ranges are marked 
underneath the spectra for the measured line emission (from left to right 
\Hgamma, \HeII, \Hbeta, $F_{\lambda}$(5177\,\AA), \HeI, and \Halpha). The 
vertical scale is in 
units of $10^{-16}$\,ergs\ s$^{-1}$\,cm$^{-2}$\,\AA$^{-1}$.
Note the dramatic variability of the ``small blue bump'', i.e., the very 
broad feature that dominates the spectrum shortward of about 4200\,\AA.
At the bottom of the figure the difference spectrum corresponding to
the low and high state is shown as thick dotted line.}

\figcaption[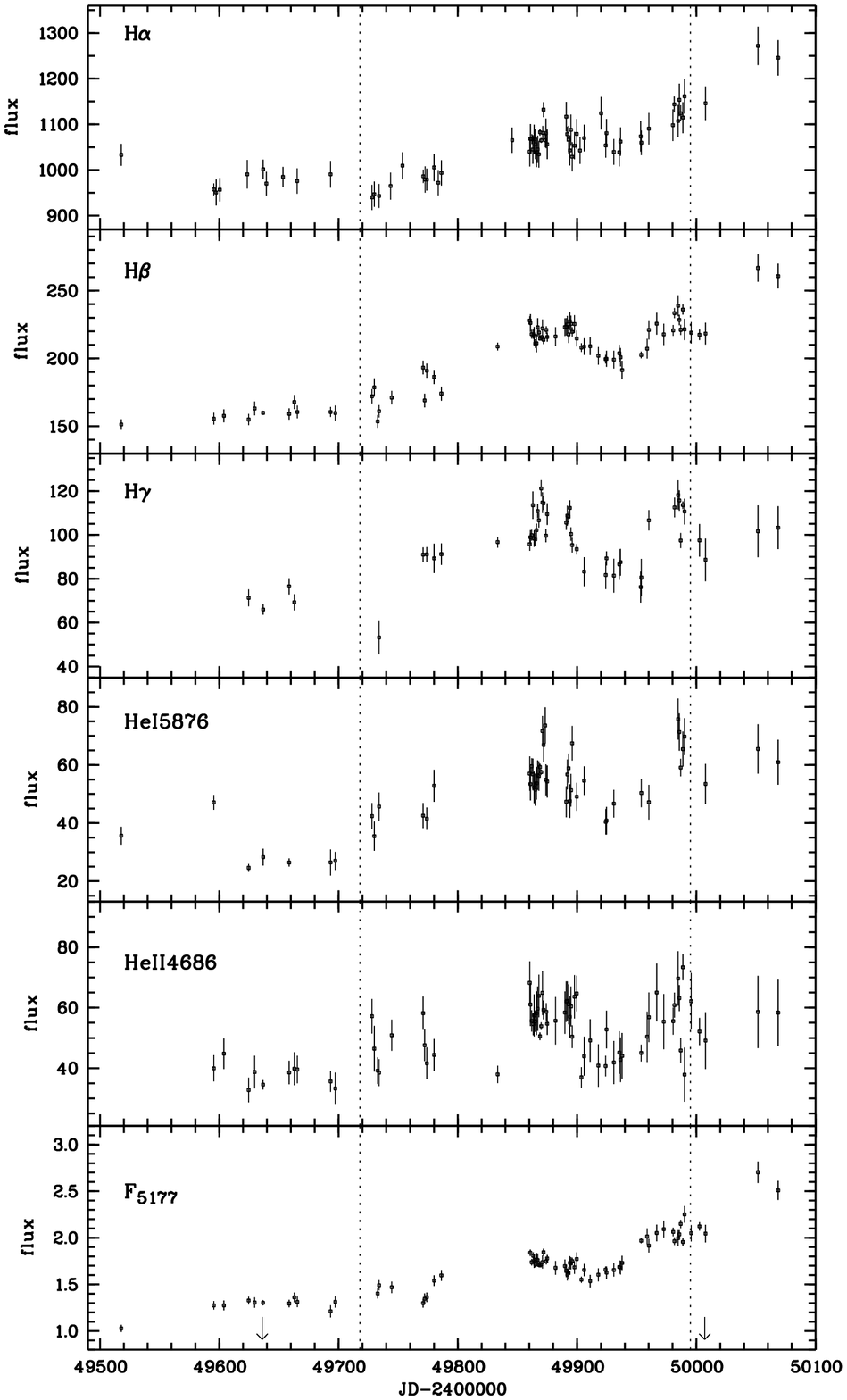]{Light curves, as available on the WWW at
the URL given above as Table 9, for the 
emission lines \Halpha, \Hbeta, \Hgamma, \HeI, and \HeII\ and
the optical continuum flux $F_{\lambda}$(5177\,\AA).
The vertical scale is in units of
$10^{-15}$\,ergs\ s$^{-1}$\,cm$^{-2}$ for the lines and
$10^{-15}$\,ergs\ s$^{-1}$\,cm$^{-2}$\,\AA$^{-1}$ for the continuum.
The dashed vertical lines mark the temporal range of the X-ray observations.}

\figcaption[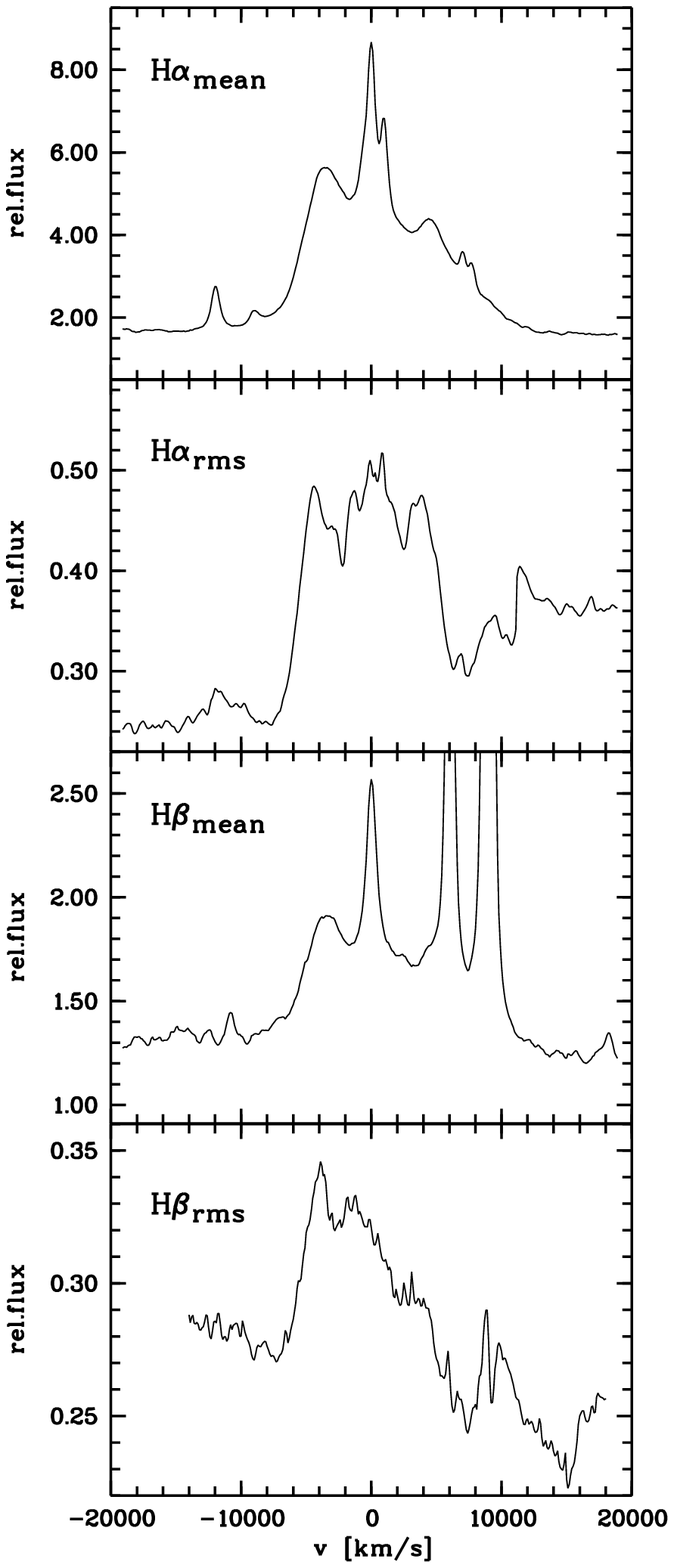]{The mean and root-mean-square (rms) spectra in the
vicinity of \Halpha\ (top) and \Hbeta\ (bottom). The vertical
scales are arbitrary. The rms spectra highlight the most variable
parts of the emission-line profiles.
In the \Halpha\ rms spectrum, narrow residuals of
[\oi]\,$\lambda6300$ are visible, and the narrow residual of
\ob\ can be seen in the \Hbeta\ rms spectrum; these are due to imperfect 
relative scaling of spectra from different sources.}

\figcaption[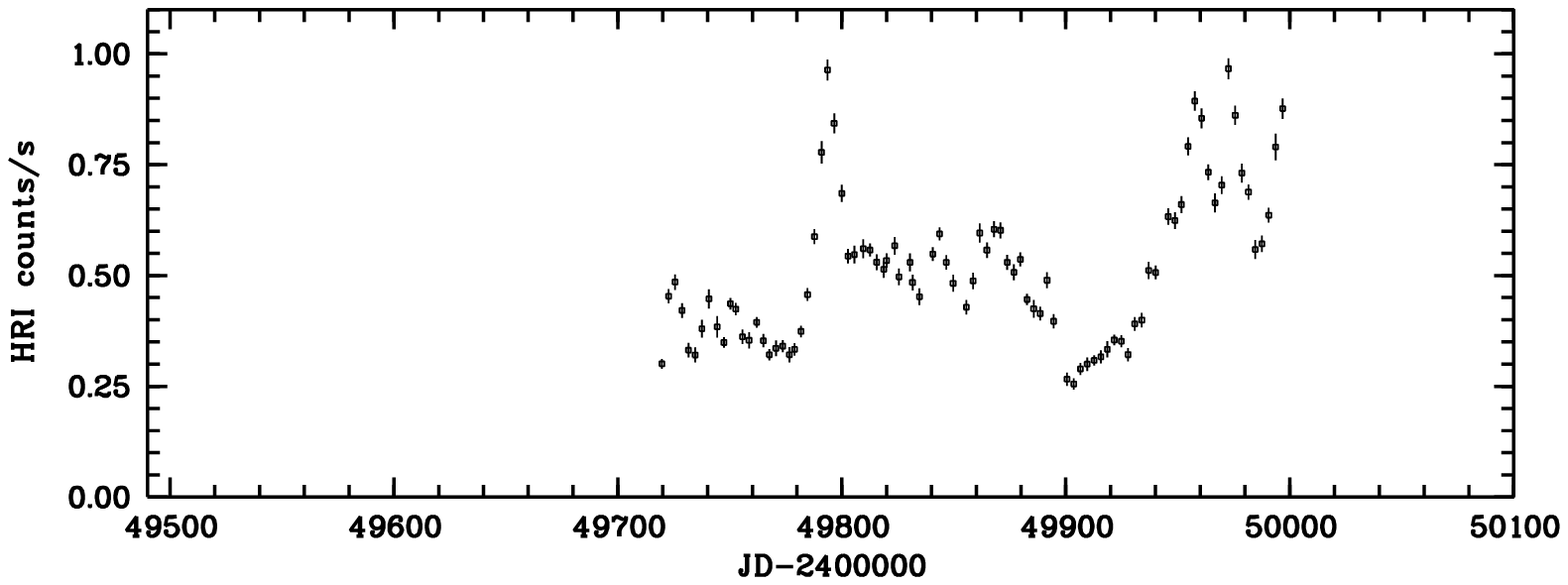]{ROSAT HRI light curve from monitoring observation 
of 3C\,390.3 (cf.\,Leighly et al.\,1997).}

\figcaption[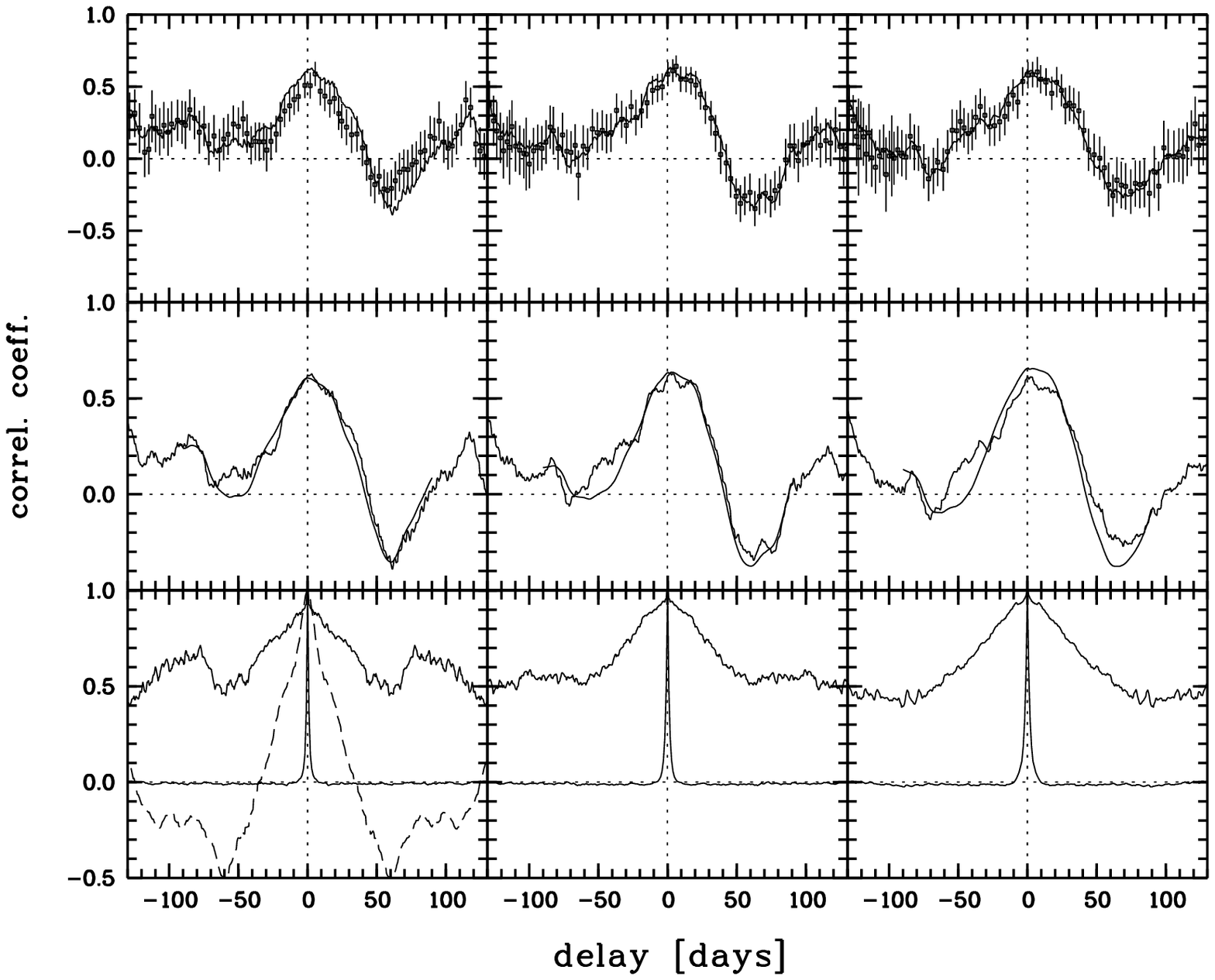]{The bottom row of panels shows the broad
autocorrelation functions (ACFs) and corresponding narrow
sampling window autocorrelation functions (ACF$_{\rm sw}$)
for the broad-band $V$, $R$, and $I$ fluxes (from left to right) available 
as Table 2 on the WWW at the URL given above. In the lower left panel the 
ACF of the X-ray variations is shown as long-dashed line for comparison with 
the ACFs of the optical variations.
In the middle row of panels the ICCF is displayed together with the 
corresponding mean ICCF which is calculated within the routine
to compute the cross-correlation peak distribution (CCPD) 
(cf.\,Maoz \& Netzer 1989). The smooth curve of the simulated mean ICCF
and the ICCF of the real observations are very similar.
The top row of panels shows the results of cross-correlating each of 
these light curves with the simultaneous X-ray light curve from Leighly 
et al.\ (1997). The ICCF is shown as a connected line, and the ZDCF is 
shown as individual points with associated error bars.}

\figcaption[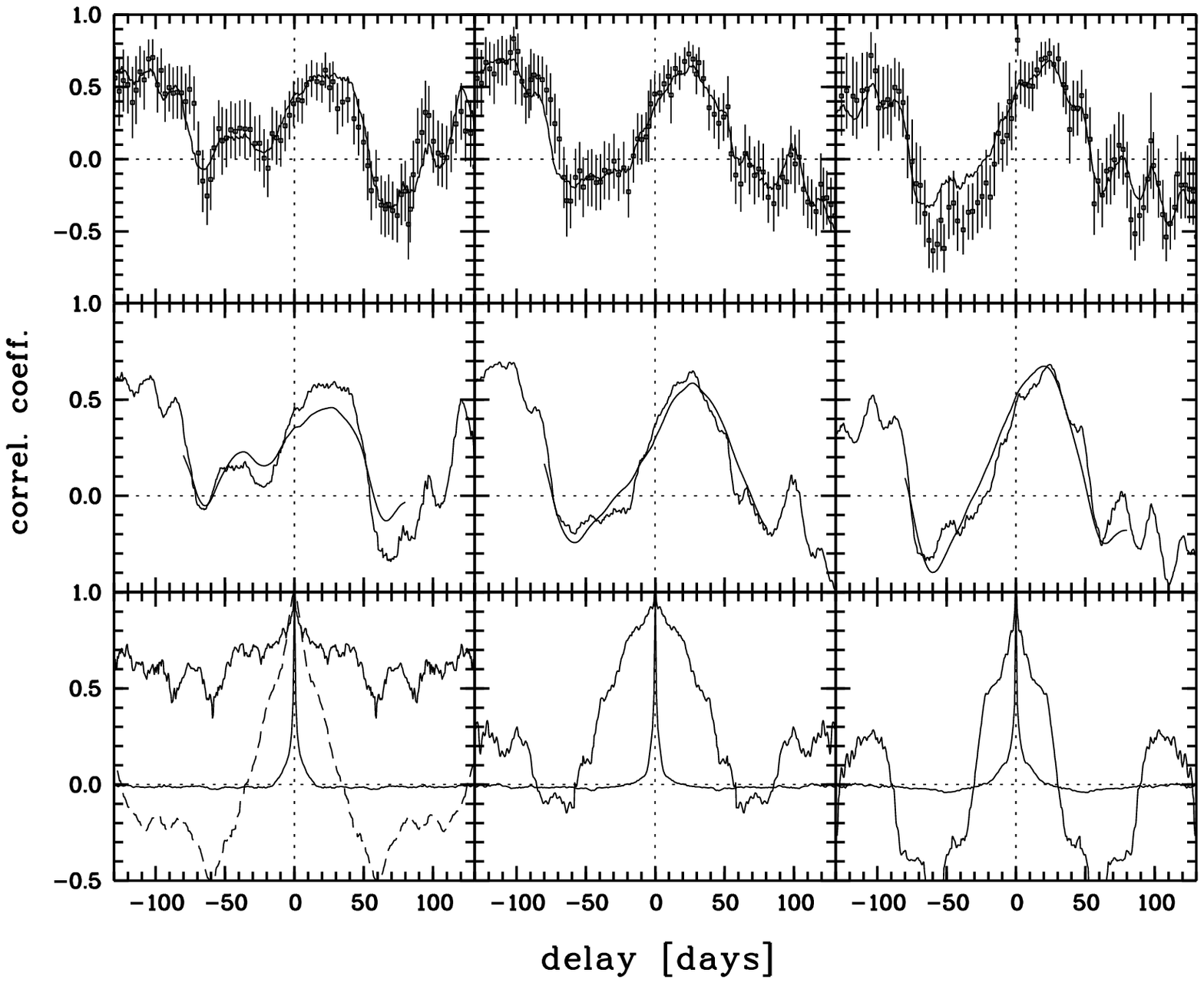]{Cross-correlation functions for the broad Balmer emission 
lines \Halpha, \Hbeta, and \Hgamma\, (from left to right). The 
cross-correlation functions are plotted as in Fig.\ 7.}

\figcaption[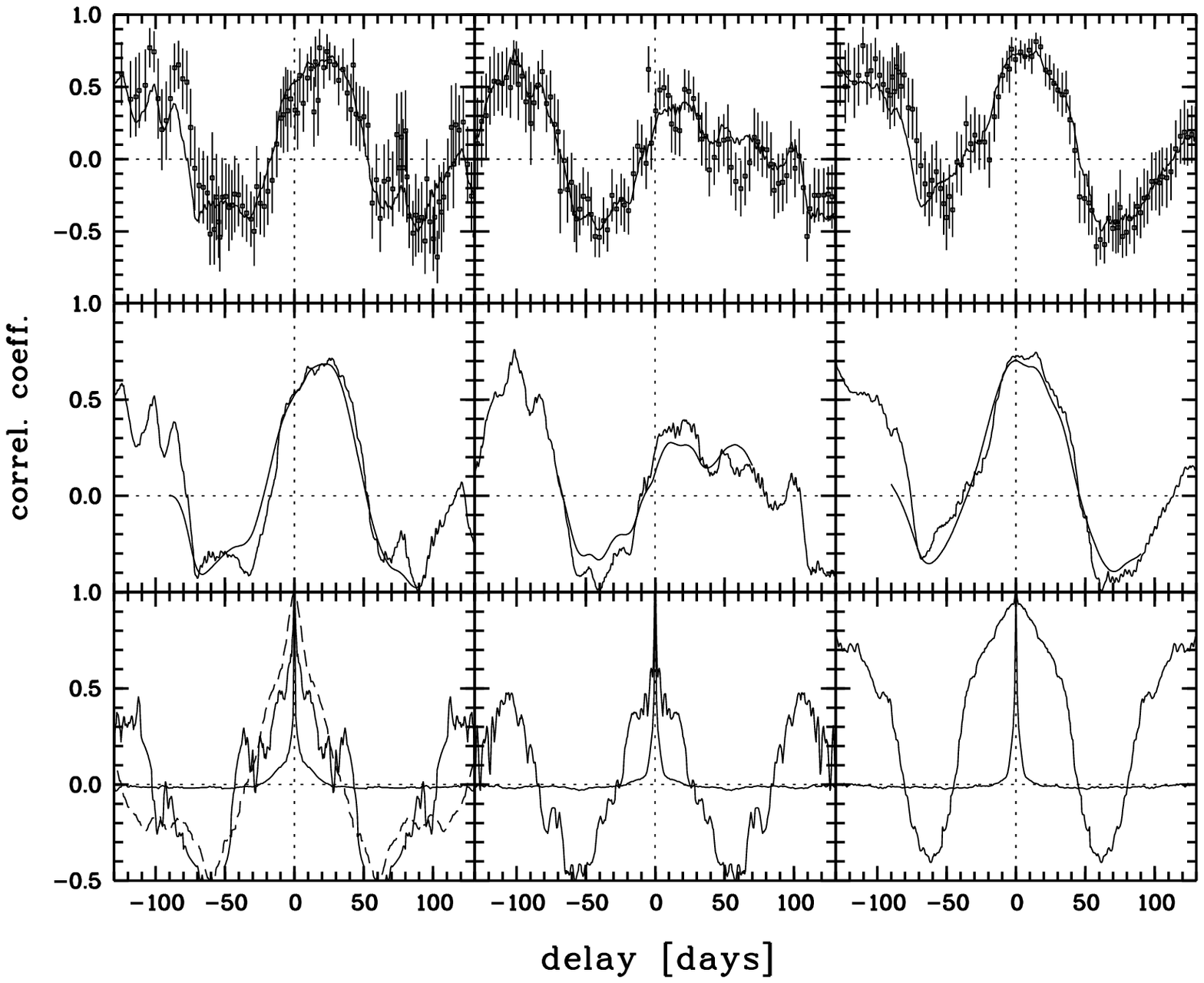]{Cross-correlation functions for the \HeI\ and \HeII\
emission lines, and for the spectrophotometric optical continuum
flux $F_{\lambda}$(5177\,\AA) (from left to right). The cross-correlation 
functions are plotted as in Fig.\ 7.}

\figcaption[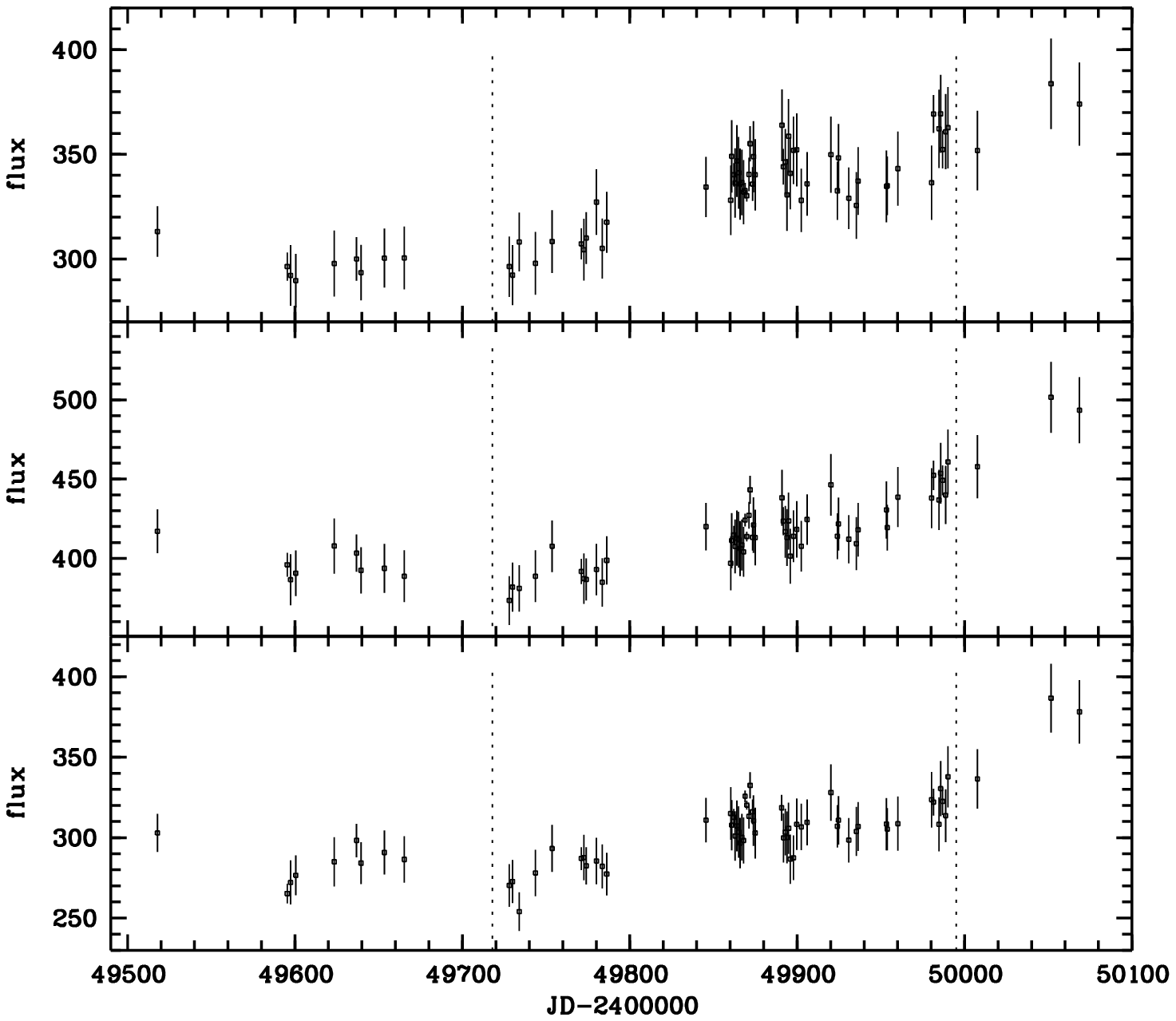]{Light curves for each of the parts of the \Halpha  
 emission line profile. The light curve of the blue wing is shown in the 
 top panel, the middle panel displays the light curve of the line core, and 
 at the bottom the light curve of the red wing is presented. The vertical 
 scale is in units of $10^{-15}$\,ergs\ s$^{-1}$\,cm$^{-2}$. The dashed 
 vertical lines mark the temporal range of the X-ray observations.}

\figcaption[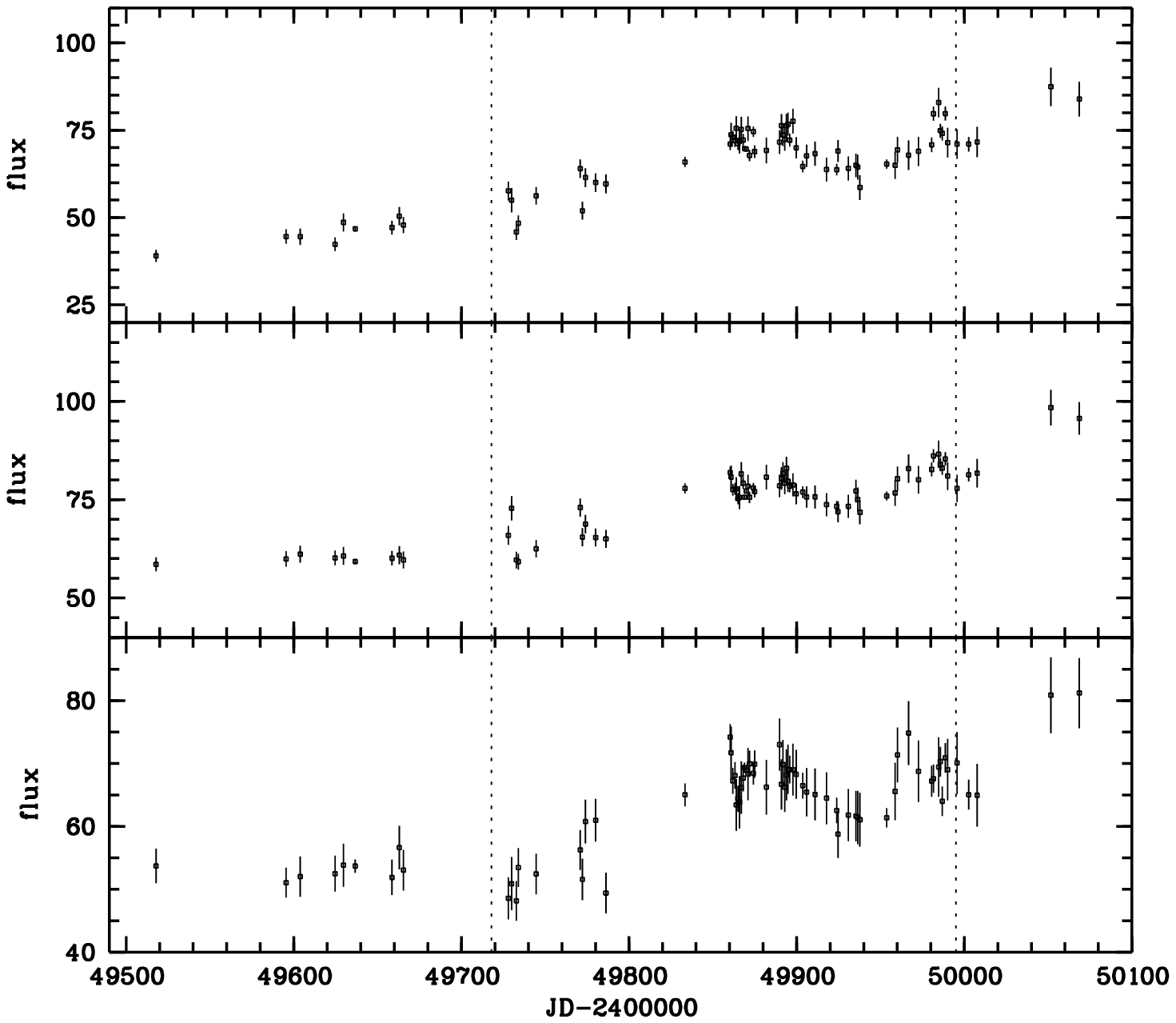]{Light curves for each of the parts of the \Hbeta 
 emission line profile. The light curve of the blue wing is shown in the top 
 panel, the middle panel displays the light curve of the line core, and at 
 the bottom the light curve of the red wing is presented. The vertical scale 
 is in units of $10^{-15}$\,ergs\ s$^{-1}$\,cm$^{-2}$. The dashed vertical 
 lines mark the temporal range of the X-ray observations.}

\figcaption[f12]{The top row of panels shows the results of cross-correlating 
the blue wing (left column) and red wing (right column) of the \Halpha\, 
emission line with the line core. The ICCF is shown as a connected line and 
the ZDCF is shown as individual points with associated error bars. 
In the middle row of panels the ICCFs of the profile sections are displayed 
together with the corresponding mean ICCF which is yielded by calculating
the CCPD (cf.\, Maoz \& Netzer 1989).
The corresponding autocorrelation functions are displayed in the bottom 
row of panels.}

\figcaption[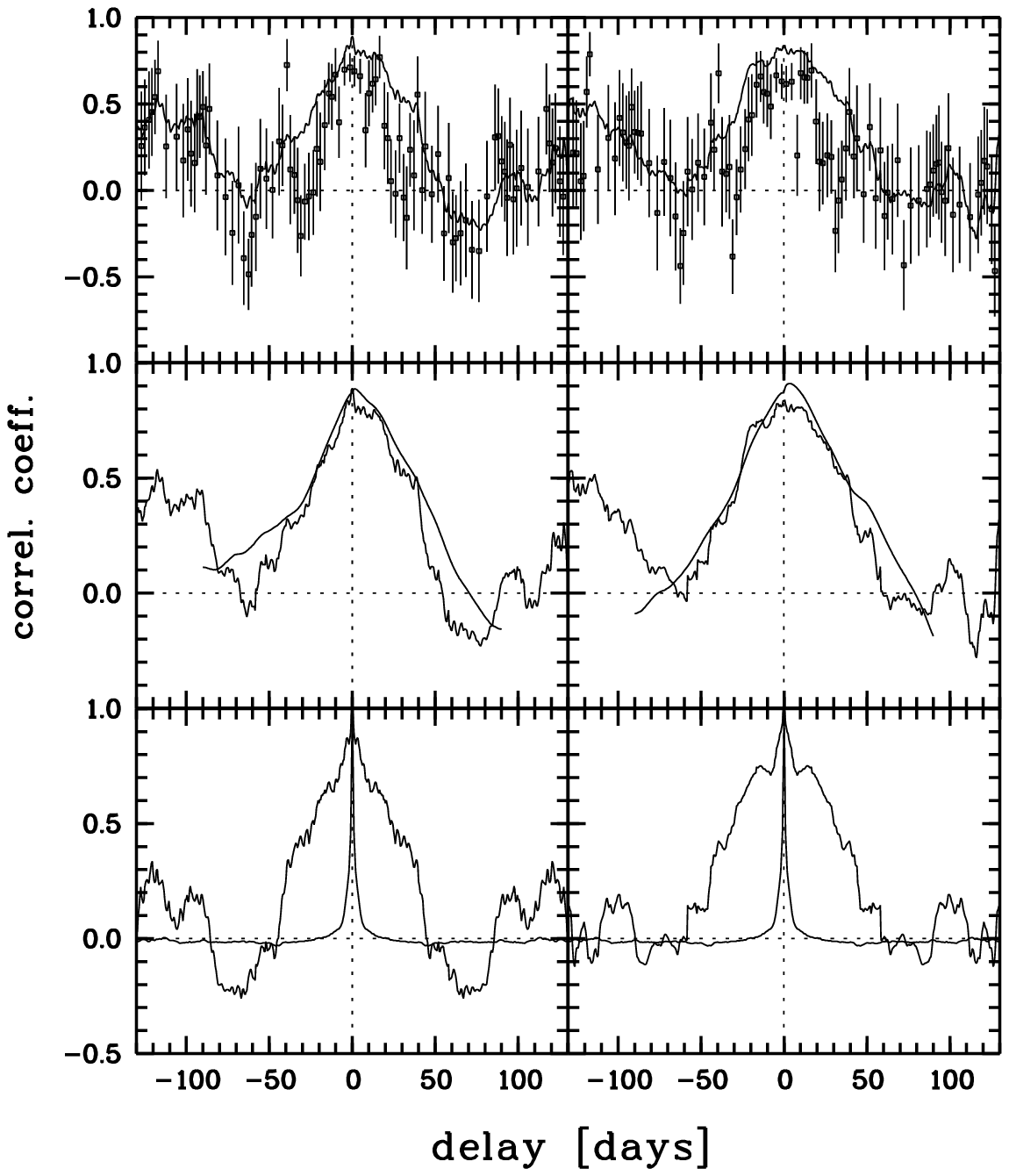]{Cross-correlation functions of the broad \Hbeta\, 
emission line sections. The results are plotted as in Fig.\,12.}

\newpage


\begin{deluxetable}{lccccccc}
\tablewidth{0pt}
\tablecaption{Overview of Observations}
\tablehead{
\colhead{ } &
\colhead{ } &
\colhead{Tel.} &
\multicolumn{4}{c}{Photometry Aperture\tablenotemark{a}} &
\colhead{Spectroscopy} \nl
\colhead{Source} &
\colhead{Code} &
\colhead{[m]} &
\colhead{$B$} &    
\colhead{$V$} &
\colhead{$R$} &
\colhead{$I$} &
\colhead{Aperture\tablenotemark{a}}\nl
\colhead{(1)} &
\colhead{(2)} &
\colhead{(3)} &
\colhead{(4)} &
\colhead{(5)} &
\colhead{(6)} &
\colhead{(7)} &
\colhead{(8)} 
}
\startdata
Perkins reflector, Lowell Obs.\tablenotemark{b} &A&1.8&---&---&---&---&5x7.5\nl
Shane reflector, Lick Obs.\tablenotemark{b}&B&3.0&---&---&---&---&4x10,2x5.6\nl
Mt.\ Hopkins Observatory       &C & 1.6&---&---&---&---&3x3,2x3,1x3\nl
Shajn reflector, Crimean Obs.  &D & 2.6&---&---&---&---&3x11\nl
Beijing Observatory	       &E & 2.2&---&---&---&--- &4x7.2, 4x8.4,\nl
                               &  &     &   &   &   &    &4x12, 4x10.3\nl
                               &  &     &   &   &   &    &4x3.6, 2.5x9.6 \nl
                               &  &     &   &   &   &    &3.5x9.6\nl
Steward Observatory            &F&2.3&2.5-3.3\tablenotemark{c}
                        &2.5-3.3\tablenotemark{c} 
			&2.5-3.3\tablenotemark{c}&---&2.5x12\nl
Calar Alto Observatory         &G1& 1.2&---&---&25.0&25.0&---\nl
Calar Alto Observatory         &G2& 2.2&14.0&14.0&14.0&14.0&2x14.1,2x10\nl
MMT Observatory   	      &H & 4.5&---&---&---&---&2x10,1x7.2,2x6\nl
Isaac Newton Telescope 		&I & 2.5&---&---&---&---&1.62x6.5\nl
Special Astrophysical Obs. &J & 1.0&4.0,4.4&5.3,4.8&5.4,8.0&---&---\nl
Special Astrophysical Obs. &J & 6.0&---&---&---&---&3x3.6,3.6x3.6\nl
McDonald Observatory	      &K &2.7&---&---&---&---&2x7.2\nl
Landessternwarte Heidelberg\tablenotemark{d}&L & 0.7&---&---&15.0&15.0&---\nl
James Gregory Tel., St.\ Andrews &M & 0.9&---&47.0&47.0&47.0&---\nl
RoboScope, Indiana Univ.	& N & 0.4&---%
				&var.\tablenotemark{c}&---&---&---\nl
Wise Observatory\tablenotemark{e}&O & 1.0&10.0&10.0&12.0&12.0&---\nl
Vainu Bappu Observatory		&P1&2.3&---&7.85&---&---&---\nl
Vainu Bappu Observatory		&P2&1.0&---&6.06&---&---&---\nl
Behlen Observatory\tablenotemark{f}&Q & 0.8&---&13.7&---&---&---\nl
Center for Basement Astrophys.  &R& 0.7&--- &6.0 &6.0 &6.0 &\nl
Hoher List Obs., Bonn 		&S1& 0.6&---&3.0 &3.0&---&---\nl
Hoher List Obs., Bonn		&S2& 0.4&---&5.0 &5.0&---&---\nl
SARA Telescope, Kitt Peak 	&T & 0.9&---&---&---&20.0 &---\nl
Shanghai Observatory            &U & 1.6&---&6-7.5&6-7.5&6-7.5&---\nl
\enddata
\tablenotetext{a}{The aperture size is given in units of arcsec.}
\tablenotetext{b}{obtained spectra were used to built up the light curves}
\tablenotetext{c}{Aperture was optimized with respect to the seeing 
                  during the exposure.}
\tablenotetext{d}{obtained broad-band measurements (R,I) were used to
                  build up the light curves}
\tablenotetext{e}{obtained broad-band measurements (B) were used to
                  build up the light curves}
\tablenotetext{f}{obtained broad-band measurements (V) were used to
                  build up the light curves}
\end{deluxetable}

\clearpage

\setcounter{table}{3}

\begin{deluxetable}{lcccc}
\tablewidth{0pt}
\tablecaption{Adopted Magnitudes for Standard Stars}
\tablehead{
\colhead{Star} &
\colhead{$B$\,\tablenotemark{a}} &    
\colhead{$V$\,\tablenotemark{a}} &
\colhead{$R$\,\tablenotemark{b}} &
\colhead{$I$\,\tablenotemark{b}} 
}
\startdata
A (HST GS 4591:756)   	&12.74&11.71&12.11&11.34 \nl
B (HST GS 4591:850)  	&15.04&14.28&14.13&13.59 \nl
D (HST GS 4591:865)  	&15.40&14.65&14.42&13.90 \nl
\quad  (HST GS 4591:731)	&14.86&14.17&14.09&13.57 \nl
\enddata
\tablenotetext{a}{From Penston, Penston, \& Sandage (1971). 
Estimated uncertainty $0.02$\,mag.}
\tablenotetext{b}{From Calar Alto 2.2-m telescope observations,
1994 December. Estimated uncertainty $0.01$\,mag. ({\it R}) and 
 $0.04$\,mag.\,({\it I})}
\end{deluxetable}

\clearpage

\setcounter{table}{4}

\begin{deluxetable}{lcccc}
\tablewidth{0pt}
\tablecaption{Scaling Factors for Photometric Subsets}
\tablehead{
\colhead{ } &
\multicolumn{4}{c}{Additive Constant (magnitudes)}  \nl
\colhead{Sample} &
\colhead{$B$\,\tablenotemark{a}} &    
\colhead{$V$\,\tablenotemark{b}} &
\colhead{$R$\,\tablenotemark{c}} &
\colhead{$I$\,\tablenotemark{c}} 
}
\startdata
N &---             &$-0.207\pm0.065$& ---             &---\nl
F &$+1.029\pm0.024$&$-0.240\pm0.001$& $+2.125\pm0.013$&---\nl
G1& ---            & ---             &$ +0.007\pm0.007$&$+0.244\pm0.000$\nl
G2& ---            & ---             &$ 14.074        $&0.305\nl
J &$-0.440\pm0.000$& $-0.267\pm0.072$&$+0.080\pm0.012$ &---\nl
M &---             &$+13.499\pm0.039$&$+13.099\pm0.065$&$+13.246\pm0.033$\nl
O &---             &$ -0.022\pm0.001$&$-0.101\pm0.019$ &$+0.138\pm0.000$\nl
P1&---             &$ -0.131\pm0.075$& ---             &---\nl
P2&---             &$ -0.120\pm0.000$& ---             &---\nl
R &---             &$ -0.380\pm0.000$&$+11.070\pm0.070$&$+11.189\pm0.000$\nl
S1&---             &$ -0.230\pm0.000$&$-0.885\pm0.047$&---\nl
S2&---             &$ -0.080\pm0.000$&$-0.841\pm0.008$&---\nl
T &---             & ---             & ---             &$+0.300\pm0.000$\nl
U &---             &$+11.480\pm0.051$&$+10.939\pm0.044$&$+11.091\pm0.047$\nl
\enddata
\tablenotetext{a}{Relative to Wise Observatory subset (O).}
\tablenotetext{b}{Relative to Behlen Observatory subset (Q).}
\tablenotetext{c}{Relative to Landessternwarte Heidelberg subset (L).
                  Calibration based on observations with Calar Alto 
                  2.2-m telescope, 1994 December.}
\end{deluxetable}

\clearpage

\setcounter{table}{5}

\begin{deluxetable}{cc}
\tablewidth{0pt}
\tablecaption{Extended Source Correction Factor $G$}
\tablehead{
\colhead{Data} &
\colhead{$G$} \nl
\colhead{Set} &
\colhead{[$10^{-15}$\,ergs\ s$^{-1}$\,cm$^{-2}$\,\AA$^{-1}$]} 
}
\startdata
A       &$ 0.00$ 	\nl
B       &$ 0.79\pm0.40$ \nl
C       &$ 1.17\pm0.64$ \nl
D       &$-0.22\pm0.40$ \nl
E       &$ 1.35\pm0.93$ \nl
F       &$ 1.21$        \nl
G2      &$ 1.23$        \nl
G2      &$ 0.79\pm0.40$ \nl
H       &$ 1.57$        \nl
H       &$ 3.29\pm0.22$ \nl
I       &$ 1.58\pm0.93$ \nl
J       &$ 1.43$        \nl
K       &$ 1.09\pm0.37$ \nl
\enddata
\end{deluxetable}

\clearpage

\setcounter{table}{6}

\begin{deluxetable}
{lcclcc}
\tablewidth{0pt}
\tablecaption{Sampling Characteristics}
\tablehead{
\colhead{Feature} &
\colhead{$N$} &
\colhead{Interval} &
\colhead{Feature} &
\colhead{$N$} &
\colhead{Interval} \nl 
\colhead{} &
\colhead{} &
\colhead{[days]} &
\colhead{} &
\colhead{} &
\colhead{[days]}
} 
\startdata
$B$   & 21&15.6$\pm$17.8&H$\gamma$                  & 55& 8.2$\pm$14.0 \nl 
$V$   &244& 1.4$\pm$ 2.5&He\,{\sc ii}\,$\lambda4686$&100& 4.8$\pm$ 9.1 \nl
$R$   &206& 1.8$\pm$ 2.7&H$\beta$                   &104& 5.3$\pm$11.2 \nl
$I$   &149& 2.4$\pm$ 3.4&He\,{\sc i}\,$\lambda5876$ & 60& 9.3$\pm$17.0 \nl
$F_{\lambda}$(5177\,\AA)& 97& 5.7$\pm$12.8&H$\alpha$& 84& 6.6$\pm$15.5 \nl
\enddata
\end{deluxetable}

\clearpage

\setcounter{table}{7}

\begin{deluxetable}{lc}
\tablewidth{0pt}
\tablecaption{Integration Limits}
\tablehead{
\colhead{ } &
\colhead{Wavelength} \nl
\colhead{Feature} &
\colhead{Range [\AA ]}
}
\startdata
H$\alpha\,\lambda6563$	& 6753--7100 \nl
He\,{\sc i}\,$\lambda5876$	& 6045--6355 \nl
$F_{\lambda}$(5177\,\AA)& 5460--5470 \nl
H$\beta\,\lambda4861$   & 5003--5260 \nl
He\,{\sc ii}\,$\lambda4686$      & 4823--5012 \nl
H$\gamma\,\lambda4340$	& 4472--4701 \nl
\enddata
\end{deluxetable}

\clearpage

\setcounter{table}{9}

\begin{deluxetable}{lcc}
\tablewidth{250pt}
\tablecaption{Comparison of Uncertainty Estimates}
\tablehead{
\colhead{Feature} &
\colhead{$\sigma_{\rm est}$\,\tablenotemark{a}} &
\colhead{$\sigma_{\rm obs}$\,\tablenotemark{b}} 
}
\startdata
$B$   				& 0.036 & 0.027 \nl
$V$   				& 0.048 & 0.043 \nl
$R$   				& 0.020 & 0.015 \nl
$I$   				& 0.019 & 0.018 \nl
H$\gamma$			& 0.062 & 0.055 \nl
He\,{\sc ii}\,$\lambda4686$ 	& 0.145 & 0.137 \nl
H$\beta$ 			& 0.025 & 0.031 \nl
$F_{\lambda}$(5177\,\AA)        & 0.040 & 0.041 \nl
He\,{\sc i}\,$\lambda5876$  	& 0.106 & 0.099 \nl
H$\alpha$			& 0.028 & 0.029 \nl
\enddata
\tablenotetext{a}{Mean fractional uncertainty based on
point-to-point differences between closely spaced
(i.e., $\Delta t \leq 2$\,days) measurements.}
\tablenotetext{b}{Observational uncertainty based on
uncertainties assigned to individual points.}
\end{deluxetable}

\clearpage

\setcounter{table}{10}

\begin{deluxetable}{lcccc}
\tablewidth{0pt}
\tablecaption{Variability Statistics of the Entire Light Curves}
\tablehead{
\colhead{Feature} &
\colhead{Mean Flux\,\tablenotemark{a}} &
\colhead{RMS Flux\,\tablenotemark{a}} &
\colhead{$R_{\rm max}$} &
\colhead{$F_{\rm var}$} 
}
\startdata
$B$                             &$2.55$  &$0.36$&$1.69$&$0.124$\nl
$V$                             &$2.62$  &$0.41$&$1.98$&$0.127$\nl
$R$                             &$2.08$  &$0.21$&$1.57$&$0.083$\nl
$I$                             &$1.33$  &$0.12$&$1.49$&$0.070$\nl
H$\gamma$			&  $96.5$&$14.9$&$2.27$&$0.093$\nl
He\,{\sc ii}\,$\lambda4686$ 	&  $51.5$&$10.3$&$2.61$&$0.131$\nl
H$\beta$ 			& $206.4$&$24.6$&$1.76$&$0.088$\nl
$F_{\lambda}$(5177\,\AA)	&   $1.7$&$ 0.3$&$2.63$&$0.122$\nl
He\,{\sc i}\,$\lambda5876$  	&  $52.8$&$12.2$&$3.08$&$0.130$\nl
H$\alpha$			&$1056.0$&$64.8$&$1.35$&$0.032$\nl
\enddata
\tablenotetext{a}{Continuum flux and broad band fluxes in units of 
$10^{-15}$\,ergs\ s$^{-1}$\,cm$^{-2}$\,\AA$^{-1}$;
line fluxes in units of 
$10^{-15}$\,ergs\ s$^{-1}$\,cm$^{-2}$.}
\end{deluxetable}

\clearpage

\setcounter{table}{11}


\begin{deluxetable}
{lcccc}
\tablewidth{0pt}
\tablecaption{Sampling Characteristics \tablenotemark{a}}
\tablehead{
\colhead{} &
\colhead{} &
\multicolumn{3}{c}{FWHM}{ACF$_{\rm sw}$} \nl
\colhead{Feature} &
\colhead{$N$} &
\colhead{ACF} &
\colhead{ICCF} &
\colhead{[days]} \nl
\colhead{(1)} &
\colhead{(2)} &
\colhead{(3)} &
\colhead{(4)} &
\colhead{(5)} 
}
\startdata
$V$                        &126& 89& 53& 1.5\nl
$R$                        &104&121& 50& 2.0\nl
$I$                        & 82&132& 53& 2.5\nl
$F_{\lambda}$(5177\,\AA)   & 55& 69& 53& 1.8\nl
H$\gamma$                  & 42& 36& 49& 1.7\nl 
He\,{\sc ii}\,$\lambda4686$& 58&  8& 33& 1.8\nl
H$\beta$                   & 59& 64& 52& 1.8\nl
He\,{\sc i}\,$\lambda5876$ & 40&  6& 55& 1.5\nl
H$\alpha$                  & 54&105& 56& 1.7\nl
\enddata
\tablenotetext{a}{X-ray restricted and rebinned light curves.}
\end{deluxetable}

\clearpage

\setcounter{table}{12}

\begin{deluxetable}{lccccccccc}
\tablewidth{0pt}
\tablecaption{Cross-Correlation Results \tablenotemark{a}}
\tablehead{
\colhead{} &
\colhead{} &
\multicolumn{3}{c}{$\tau_{\rm max}$ [days]} &
\colhead{$\Delta\tau_{\rm max}$} &
\multicolumn{3}{c}{$\tau_{\rm cent}$ [days]} &
\colhead{$\Delta\tau_{\rm cent}$} \nl
\colhead{Feature} &
\colhead{$r_{\rm max}$} &
\colhead{ICCF} &
\colhead{ZDCF} &
\colhead{DCF}  &
\colhead{(days)}&
\colhead{ICCF} &
\colhead{ZDCF} &
\colhead{DCF}  &
\colhead{(days)} \nl
\colhead{(1)} &
\colhead{(2)} &
\colhead{(3)} &
\colhead{(4)} &
\colhead{(5)} &
\colhead{(6)} &
\colhead{(7)} &
\colhead{(8)} &
\colhead{(9)} &
\colhead{(10)}
}
\startdata
$V$                        &0.62& 3.5  & 5.3& 4.9&4.5 & 3.5& 5.0& 5.0&3.0 \nl
$R$                        &0.63& 3.5  & 4.8& 3.8&5.6 & 5.5& 5.8& 4.0&3.9\nl
$I$                        &0.61& 3.5  & 4.5& 3.7&6.2 & 7.9& 7.0& 4.7&4.2 \nl
$F_{\lambda}$(5177\,\AA)\tablenotemark{b}
			   &0.73&$-2.5$& 8.3&11.4&6.2 & 5.3& 8.5& 8.5&2.8 \nl
H$\gamma$                  &0.66& 24.5 &23.7&24.1&6.2 &16.8&13.8&20.8&5.5 \nl
He\,{\sc ii}\,$\lambda4686$&0.37& 13.0 & 5.9&10.4&8.3 & 8.6& 4.6&17.8&7.8 \nl
H$\beta$                   &0.64& 24.5 &24.4&12.5&5.3 &22.9&23.2&15.4&3.9 \nl
He\,{\sc i}\,$\lambda5876$ &0.69& 25.0 &23.4&11.4&9.5 &18.6&22.6&17.2&8.5 \nl
H$\alpha$                  &0.53& 29.0 &18.7&11.2&10.3 &19.3&20.6&15.9&9.5 \nl
\enddata
\tablenotetext{a}{Light curves restricted approximately to the 
period of X-ray monitoring, JD2449719--JD2449997.}
\tablenotetext{b}{Light curves restricted approximately to the 
period of overlap, JD2449732--JD2449996.}

\end{deluxetable}

\clearpage

\setcounter{table}{14}

\begin{deluxetable}{llccccccccc}
\tablewidth{0pt}
\tablecaption{Cross-Correlation Results for Line 
Profile Sections\tablenotemark{a}}
\tablehead{
\colhead{First} &
\colhead{Second} &
\colhead{} &
\multicolumn{3}{c}{$\tau_{\rm max}$ [days]} &
\colhead{$\Delta\tau_{\rm max}$} &
\multicolumn{3}{c}{$\tau_{\rm cent}$ [days]} &
\colhead{$\Delta\tau_{\rm cent}$} \nl
\colhead{Series} &
\colhead{Series} &
\colhead{$r_{\rm max}$} &
\colhead{ICCF} &
\colhead{ZDCF} &
\colhead{DCF}  &
\colhead{[days]}&
\colhead{ICCF} &
\colhead{ZDCF} &
\colhead{DCF}  &
\colhead{[days]} 
}
\startdata
H$\beta _{\rm core}$ & H$\beta _{\rm blue\,wing}$ 
   &0.89& 0.0  & 1.0 &  -4.0 & 5.7 & 5.6& 3.1 & 0.0 & 5.6 \nl
H$\beta _{\rm core}$ & H$\beta _{\rm red\,wing}$
   &0.84& 0.5  & -1.2 &  0.0 & 8.6 & 0.3& 0.4 & 8.8 & 8.7 \nl
H$\alpha _{\rm core}$ & H$\alpha _{\rm blue\,wing}$ 
   &0.86& 0.0  & 4.4 &  -4.0 & 18.4 & 3.9& 2.1 & -4.0 & 17.8 \nl
H$\alpha _{\rm core}$ & H$\alpha _{\rm red\,wing}$   
   &0.87& 0.0  & 1.2 &  0.0 & 22.6 &$-0.5$& -0.9 & -1.9 & 19.0 \nl
\enddata
\tablenotetext{a}{Light curves restricted approximately to the 
period of X-ray monitoring, JD2449719--JD2449997.}
\end{deluxetable}

\end{document}